\documentclass[11pt]{article}
\usepackage{amsmath,amssymb,bm,epsf,epsfig,graphicx}
\input epsf.sty
\usepackage[utf8]{inputenc}
\usepackage[english]{babel}
\usepackage{amsmath}
\usepackage{latexsym} 
\usepackage{amsfonts}
\usepackage{mathtools}
\usepackage{amssymb}
\usepackage[left=3cm,right=3cm,top=3.1cm,bottom=3.1cm]{geometry}
\usepackage{cite}
\usepackage{color}

\usepackage{hyperref}
\numberwithin{equation}{section}

\newcommand{\Tr}{\mathop{\rm Tr}\nolimits}

\def\bra#1{\langle #1 |}

\let\eps = \epsilon

\def \be {\begin{eqnarray}}
\def \ee {\end{eqnarray}}
\def \bdm {\begin{displaymath}}
\def \edm {\end{displaymath}}

\def \tr{{\rm tr}}

\def\del {\partial}
\def\0{\nonumber}

\def \VV {{\mathbb V}}

\def \HH {{\mathbb H}}

\def\N{\mathcal {N}}

\begin{document}

\vspace*{3.1cm}

\centerline{\Large \bf Localization of effective actions in }\vspace{.4cm}
\centerline{\Large \bf  open superstring field theory: } \vspace{.4cm}
\centerline{ \large\bf  small Hilbert space}\vspace{.2cm}

\vspace*{.1cm}

\begin{center}

{\large Carlo Maccaferri\footnote{Email: maccafer at gmail.com} and Alberto Merlano\footnote{Email: albemerlano at gmail.com} } 
\vskip 1 cm
{\it Dipartimento di Fisica, Universit\'a di Torino, \\INFN  Sezione di Torino and Arnold-Regge Center\\
Via Pietro Giuria 1, I-10125 Torino, Italy}
\end{center}

\vspace*{6.0ex}

\centerline{\bf Abstract}
\bigskip
We consider the algebraic effective couplings for open superstring massless modes in the framework of the $A_\infty$ theory in the small Hilbert space. Focussing on quartic  couplings, we reduce the effective action of the $A_\infty$ theory to the Berkovits one where we have already shown that such couplings are fully computed from contributions at the boundary of moduli space, when the massless fields are appropriately charged under an $\N=2$ $R$-symmetry. Here we offer a  proof of localization which is  in the small Hilbert space. Our analysis shows that the flat directions of the quartic potential are controlled by ADHM-like constraints, which are directly related to the localization channels of the effective action. In particular we  give evidence for the existence of exactly marginal deformations corresponding to blowing up moduli in the $D(p+4)/Dp$ system in the framework of string field theory. \vfill \eject

\baselineskip=16pt

\tableofcontents
\newpage
\section{Introduction}\label{intro}
Complete constructions of superstring field theories have been  achieved in the last years and we have now at our disposal consistent actions for both open and closed superstrings in either small or large Hilbert space formalisms \cite{Kunitomo:2019glq, Erler:2017onq, Konopka:2016grr, Erler:2016ybs, Sen:2015uaa,  Kunitomo:2015usa} . So we are now in a stage where we could possibly use these actions for concrete calculations. However the non-polinomiality that is inherent to both open and closed superstrings makes it difficult to perform typical QFT calculations such as effective actions, mass renormalization or related observables.
In this regard it is  useful to explore whether  questions involving physical quantities can  possibly be answered  by-passing the complicated off-shell structure of the fundamental string field theories.

As a  step in this direction, in \cite{loc1} we revisited the construction of the tree-level effective action for open (NS) superstrings, integrating out massive fields from the Berkovits WZW-like theory \cite{Berkovits}   as was originally done in \cite{BS}. Our analysis surprisingly showed that the whole contribution to the effective action up to quartic order is entirely captured by singular naively-vanishing terms at the boundary of moduli space of four-punctured disks. These terms arise by carefully accounting for the failure of the propagator to truly invert the BRST charge
\be
\left[Q_B,\frac{b_0}{L_0}\right]=1-P_0,\label{QP}
\ee
where $P_0$ is the projector on the kernel of $L_0$, which is responsible for the contributions at the boundary of moduli space. In particular when we  compute the effective action of massless fields that are charged under the $U(1)$ $R$-symmetry of the underlying  $\N=2$ SCFT describing the superstring background we consider, the whole quartic potential vanishes {\it except} for ``singular''  terms containing $P_0$. 
This has the consequence that such quartic couplings are in fact computed by two-point functions of so-called {\it auxiliary fields}, which are  given by the projection to level zero of the star product (or OPE) of the physical fields. For generic momenta these contributions are vanishing but at zero momentum (where the algebraic couplings live) they play an essential role.  In particular, after computing the ghost correlators, our result for the four-point coupling of massless open superstring fields of the WZW theory in the large Hilbert space
\be
\Phi_A=\Phi_A^{+}+\Phi_A^{-}=c\gamma^{-1}\left(\VV_{\frac12}^{(+)}+\VV_{\frac12}^{(-)}\right),\label{large-split}
\ee
 boiled down to a simple combination of matter two-point functions
\be
S_{{\rm eff},\,WZW}^{(4)}(\Phi_A)=\tr\left[\bra{\HH_{1}^{(+)}}\,\HH_1^{(-)}\rangle + \frac14\bra{\HH_{0}}\,\HH_0\rangle \right]\label{seffmatt}.
\ee
In the above expressions  ``tr'' is the trace in Chan-Paton space and  $\HH_1^{(\pm)}$ and $\HH_0$  are the leading contributions of the OPE of matter physical superconformal primaries (including space-time polarization) at zero momentum, $\VV_{\frac12}^{(\pm)}$, which have been decomposed in $J=\pm1$ eigenstates of a $U(1)$ $R$-charge in a convenient $\N=2$ organization of the matter SCFT
\be
\HH_1^{(\pm)}(x)&=&\lim_{\eps\to0}\VV_{\frac12}^{(\pm)}(x+\eps)\VV_{\frac12}^{(\pm)}(x-\eps),\label{H1}\\
\HH_0(x)&=&\lim_{\eps\to0}\, (2\eps)\, \left(\VV_{\frac12}^{(+)}(x+\eps)\VV_{\frac12}^{(-)}(x-\eps)-\VV_{\frac12}^{(-)}(x+\eps)\VV_{\frac12}^{(+)}(x-\eps)\right).\label{H2}
\ee
It is the scope of the present paper to obtain the same result in the context of the $A_\infty$ theory in the small Hilbert space by Erler, Konopka and Sachs \cite{EKS}. 
In a sense, this is obviously expected since the two theories are known to be related by partial gauge fixing plus field redefinition \cite{Erler:2015rra, Erler:2015uoa, Erler:2015uba}. However, as already mentioned,  when we  work at zero momentum  contributions at the boundary of moduli space which arise by manipulations involving (\ref{QP}) are important (the effective action itself is generated by these contributions). So, in showing that the two effective actions give the same result, we should also carefully take into account potentially anomalous terms involving $P_0$ and we have to show that such anomalous terms are canceling.  As we will see, the vanishing of the potential anomalies that could arise in relating the Berkovits and the $A_\infty$ effective actions is a consequence of a projector condition that holds at zero momentum in the NS sector\footnote{A corresponding projector condition is at work in the Berkovits theory \cite{loc1} and reads $$P_0[\eta_0\Phi_A,Q_B\Phi_A]=0,$$ for a physical field  $\Phi_A=c\gamma^{-1}\VV_{\frac12}$.}
\be
P_0 M_2(\Psi_A,\Psi_A)=0,\label{proj-cond}
\ee
when $\Psi_A$ is a physical field in the small Hilbert space in  the form
\be
\Psi_A=c \,\VV_{\frac12}\delta(\gamma)=\eta_0\Phi_A.\label{PsiA}
\ee
Then, thanks to  \eqref{proj-cond}, the effective actions are the same in the sense that
\be
S^{(4)}_{{\rm eff}, WZW}(\xi_0\Psi_A)=S^{(4)}_{{\rm eff}, A_\infty}(\Psi_A).
\ee
Therefore if the WZW effective action localizes at the boundary of moduli space \cite{loc1} so it is for the $A_\infty$. 
When, parallel to \eqref{large-split},  we can decompose the $A_\infty$ physical field in $R$-charged eigenstates
\be
\Psi_A=\Psi^{+}+\Psi^{-}=c\left(\VV_{\frac12}^{(+)}+\VV_{\frac12}^{(-)}\right)\,\delta(\gamma),\label{small-split}
\ee
we are able to show that the quartic effective potential (the same for the $A_\infty$ and the WZW theory)  can be written in the small Hilbert space as \footnote{$P_0  \left[ \xi_0 \Psi^-, X_0 \Psi^- \right] $ is in the small Hilbert space because $P_0 \left[ \Psi^-, X_0 \Psi^- \right]=P_0 \left[ \eta_0\Phi^-, Q_B \Phi^- \right]=0$ and the same applies to $(+\leftrightarrow -)$.}
\be
S^{(4)}_{\rm eff}(\Psi_A)&=&- \dfrac{1}{8} \text{Tr}_S {\Bigg[ }[ \Psi^+ , \Psi^+ ]  P_0  \left[ \xi_0 \Psi^-, X_0 \Psi^- \right] {\Bigg] }
- \dfrac{1}{8} \text{Tr}_S {\Bigg[ }[ \Psi^- , \Psi^- ]  P_0  \left[ \xi_0 \Psi^+, X_0 \Psi^+ \right] {\Bigg] }\0\\
&&+ \dfrac{1}{8} \, \text{Tr}_S \Bigg{[}\Big{(}\left[\Psi^+, X_0 \Psi^- \right] - \left[ \Psi^-, X_0 \Psi^+ \right]\Big{)} \, P_0 \, \Big{(}\left[ \Psi^-, \xi_0 \Psi^+ \right]  + \left[\xi_0 \Psi^-, \Psi^+ \right] \Big{)} \Bigg{]}\0\\
&=&\tr\left[\bra{\HH_{1}^{(+)}}\,\HH_1^{(-)}\rangle + \frac14\bra{\HH_{0}}\,\HH_0\rangle \right].
\ee
Therefore, to quartic order, there is a unique effective potential for the zero momentum components of the massless open string fields and this potential is localized at the boundary of moduli space if the massless fields are $R$-charged.

The paper is organized as follows. In section \ref{sec:2} we review the basic aspects of the $A_\infty$ construction and we explicitly construct the tree-level effective action for massless fields by perturbatively solving the equations of motion for the heavy fields in terms of the massless fields. The effective action is expressed in terms of the $A_\infty$ multi-string vertices $M_n$ and the Siegel gauge propagator $\frac{b_0}{L_0}$. In section \ref{sec:3}, concentrating on the quartic terms, we show that, when the entries are physical, the $A_\infty$ 2-products and the products appearing in the Berkovits theory simply differ by an exact term in the small Hilbert space. This is used in section \ref{sec:4} to show that (also thank to the projector condition (\ref{proj-cond})) the effective actions derived from the WZW theory and the $A_\infty$ are identical to quartic order. As a by product of our proof, we can write the effective action of the WZW theory as a trace in the small Hilbert space.  In section \ref{sec:5}, we give a new proof that the effective action localizes at the boundary of moduli space by remaining in the small Hilbert space and decomposing the massless string field in the eigenstates of the world-sheet $\N=2$ $R$-charge $J$ in the matter sector. Given the universal form of the quartic potential, in section \ref{sec:6} we identify generic ADHM-like constraints as sufficient conditions for  flat directions. In the case of a $Dp$-$D(p-4)$ system we recover  the familiar three ADHM constraints, plus other three dual constraints which switch on a VEV for the zero momentum gauge field. We conclude with a final summary and possible further directions.
Appendix \ref{app:A}  contains the computations needed to prove \eqref{cont-equa}.
Appendix \ref{app:B} is a simple but useful comparison between our results and the reported obstruction in \cite{ivo}:  by relating the effective action at quartic order to the contraction of the interacting part of the third order equation of motion with the first order solution, we show that there is a clear and concrete disagreement between the final results of \cite{ivo} and our computations here and in \cite{loc1}.
\section{The  $A_{\infty}$ theory}\label{sec:2}

The Erler-Konopka-Sachs (EKS) action \cite{EKS} is given by the following infinite series\footnote{We normalize the vacua as in \cite{loc1} $\langle \xi(z) c \partial c \partial^2 c (w) \exp^{-2 \phi(y)} \rangle_L = \langle  c \partial c \partial^2 c (w) \exp^{-2 \phi(y)} \rangle_S =-2$. This is opposite to the convention used in \cite{BS} and as such our effective action in \cite{loc1} and here differ from \cite{BS} by an overall minus sign. To be consistent with our conventions we define the EKS action \eqref{Ainf-act} with a minus in front.}
\begin{equation}
S_{A_\infty}(\Psi) = -\sum_{n=1}^{+ \infty} \dfrac{1}{n+1} \, \omega_S \left( \Psi \, , M_n(\Psi^n) \right) .\label{Ainf-act}
\end{equation}
To start with,  $\omega_S: \mathcal{H_S} \otimes \mathcal{H_S} \longrightarrow \mathbb{C}$ is the symplectic form on the small Hilbert space. It is constructed from the BPZ inner product in the small Hilbert space
\begin{equation}
\omega_S (A,B) = (-1)^{Deg(A)} \langle A, B\rangle_S \qquad \qquad \forall \, A,B
\end{equation}
The small Hilbert space is endowed with a natural degree different from the Grassmanality but related to it
\begin{equation}
Deg(A) = A + 1, \qquad \qquad \forall A \label{grad}
\end{equation}
\begin{equation}
Deg(A \ast B) = A + B + 1, \qquad \qquad \forall A,B.
\end{equation}
The dynamical string field $\Psi$ lives in the small Hilbert space in the NS sector
\begin{equation}
 \eta_0 \Psi = 0 ,
\end{equation}
and is taken to be degree even (that is  Grassmann odd) with ghost number $1$ and picture number $-1$.

The multi-string products $$M_n: \mathcal{H}_S^{\otimes n} \rightarrow \mathcal{H}_S$$ are linear operators from the $n-$fold tensor product of the small Hilbert space back to a single copy in the small Hilbert space, which are  therefore annihilated by $\eta_0$
\begin{equation}
\left[ \eta_0 , M_n \right] = 0.
\end{equation}
The first product $M_1$ is 
\begin{equation}
M_1 = Q_B.
\end{equation} 
The consistency of the theory is realized when these multi-string products  satisfy the so-called $A_{\infty}$ relations. We list only the first three relations, which are the ones we need in this paper
\begin{equation}
0 = Q_B^2 A, \label{ainf1}
\end{equation}
\begin{equation}
0 = Q_B M_2(A,B) + M_2(Q_B A, B) + (-1)^{Deg(A)} M_2 (A, Q_B B), \label{ainf2}
\end{equation}
\begin{eqnarray}
0 &=& M_2\left( M_2(A,B), C\right) + (-1)^{Deg(A)} M_2\left(A, M_2(B,C)\right) + Q_B M_3 (A, B, C) + \nonumber \\
 && + M_3\left( Q_B A,\, B, \, C \right) + (-1)^{Deg(A)} \, M_3\left( A,\, Q_B B, \, C \right) + \nonumber \\
&& + (-1)^{Deg(A) + Deg(B)} \,  M_3\left( A,\, B, \, Q_B C \right). \label{ainf3}
\end{eqnarray}

With the definition of the gradings \eqref{grad}, the multi-string products  appearing in the action are degree odd. The symmetry property of the symplectic form in the small Hilbert space is given as follows
\begin{equation}
\omega_S(A,B) = (-1)^{Deg(A) Deg(B) +1} \omega_S(B,A).
\end{equation}
When both string fields are degree odd, the simplectic form is symmmetric in the entries. The Leibniz rule corresponding to a degree odd derivation (for example $Q_B$) of a string field is given by
\begin{equation}
\omega_S (Q_B A, B) + (-1)^{Deg(A)} \, \omega_S ( A, Q_B B) = 0 \qquad \qquad \forall \, A,B
\end{equation} 
while the cyclicity property with respect to degree odd multi-string products is
\begin{equation}
\omega_S (M_n(\Psi_1, \Psi_2, ..., \Psi_n),\Psi_{n+1}) = - (-1)^{Deg(\Psi_1)} \omega_S (\Psi_1, M_n( \Psi_2,\Psi_3, ..., \Psi_{n+1} )) \quad \forall n. \label{cyclicity}
\end{equation}
In particular this holds for $Q_B$, $M_2$ and $M_3$ multi-string products:
\begin{equation}
\omega_S (M_2(\Psi_1, \Psi_2),\Psi_3) = - (-1)^{Deg(\Psi_1)} \omega_S (\Psi_1, M_2( \Psi_2,\Psi_3)),
\end{equation}
\begin{equation}
\omega_S (M_3(\Psi_1, \Psi_2, \Psi_3),\Psi_4) = - (-1)^{Deg(\Psi_1)} \omega_S (\Psi_1, M_3( \Psi_2,\Psi_3, \Psi_4 )).
\end{equation}
These axioms of compatibility are needed to derive the correct equations of motion
\begin{equation}
\sum_{i=1}^{+ \infty} M_n (\Psi^n) = 0 \qquad \rightarrow \qquad  Q_B \Psi + M_2( \Psi^2) + M_3( \Psi^3) + ...  = 0. \label{eqa}
\end{equation}
A concrete solution to the $A_\infty$ relation is obtained by starting (for simplicity) with the  worldsheet zero mode of the PCO:
\begin{equation}
X_0 = \left[ Q_B , \xi_0 \right] = \oint \dfrac{dz}{2 \pi i} \dfrac{1}{z} X(z) \qquad , \qquad X(z) = \left[ Q_B , \xi(z) \right]
\end{equation}
and defining
\begin{equation}
M_2 ( A,B) := \dfrac{1}{3} \left[ X_0 \, m_2 (A,B) + m_2(X_0 A, B)+ m_2(A, X_0 B)    \right] \label{M2}
\end{equation} 
\begin{eqnarray}
M_3(A,B,C) &:=& \dfrac{1}{2} \, M_2(A, \bar{M}_2 (B,C)) + \dfrac{1}{2} \, M_2( \bar{M}_2 (A,B),C) - \dfrac{1}{2} \,\bar{M}_2 (M_2(A,B),C) \nonumber \\
&&  - \dfrac{1}{2} \, (-1)^{Deg(A)} \bar{M}_2 (A, M_2(B,C)) +  \dfrac{1}{2} \, Q_B-\text{exact}
\end{eqnarray}
where the BRST exact term is added in order to get a 3$-$string product in the small Hilbert space:
\begin{eqnarray}
Q_B-\text{exact} &=& +  Q_B \bar{M}_3\left( A,\, B, \, C \right) -  \bar{M}_3\left( Q_B A,\, B, \, C \right) + \nonumber \\
&& - (-1)^{Deg(A)} \,  \bar{M}_3\left( A,\, Q_B B, \, C \right) - (-1)^{Deg(A) + Deg(B)} \,  \bar{M}_3\left( A,\, B, \, Q_B C \right). \nonumber \\
\end{eqnarray}
These multi-string products are constructed from the so-called \textit{bare} multi-string product in the small Hilbert space and the \textit{dressed} multi-string products in the large Hilbert space:
\begin{equation}
m_2 (A,B) = (-1)^{Deg(A)} A \ast B = (-1)^{Deg(A)} A \, B ,
\end{equation}
\begin{equation}
\bar{M}_2 ( A,B) := \dfrac{1}{3} \left[ \xi_0 \, m_2 (A,B)- m_2(\xi_0 A, B)-(-1)^{Deg(A)} m_2(A, \xi_0 B)    \right], 
\end{equation}
\begin{equation}
m_3(A,B,C) = \dfrac{2}{3} \, \left[ m_2 \left( A, \xi_0 \, m_2(B,C) \right) + m_2 \left( \xi_0 \, m_2 (A,B),C \right)\right],
\end{equation}
\begin{eqnarray}
\bar{M}_3 \left( A, B, C \right) &=& \dfrac{1}{4}\, \left[  \xi_0 \, m_3(A,B,C)  - m_3(\xi_0 \,A, B, C) - (-1)^{Deg(A)} m_3(A, \xi_0 \,B, C) \right.\nonumber\\
&& \left.- (-1)^{Deg(A) + Deg(B)} m_3(A, B,\xi_0 \,C) \right].
\end{eqnarray}
These products are obtained placing $\xi_0$ or the PCO $X_0$ on each external state. When acting on states in the small Hibert space we have that
\begin{equation}
M_2 = \left[Q_B, \bar{M_2} \right] ,
\end{equation}
\begin{equation}
m_2 = \left[\eta_0, \bar{M_2} \right].
\end{equation}
The $Q_B-$exact part of $M_3$ is constructed in such a way that $M_3$ lies in the small Hilbert space
\begin{equation}
\left[ \eta_0 , \, M_3 \right] = 0.
\end{equation}
This part contains the dressed $3-$string product $\bar{M}_3$ which exhibits  the most symmetric assignment of $\xi_0$ to the entries. So the action of a $\eta_0$ on the dressed product is not vanishing but it gives  the bare $m_3$ product 
\begin{equation}
m_3 = \left[\eta_0, \bar{M_3} \right].
\end{equation}
\subsection{The effective action from the $A_\infty$ theory}\label{subsec:2.1}

Since we are going to perturbatively integrate out the massive fields, we start writing the $A_\infty$ theory as
\begin{equation}
S_{A_\infty} \left( \Psi \right)= -  \dfrac{1}{2} \, \omega_S \left( \Psi, Q_B \Psi \right) - \dfrac{1}{3} \, \omega_S \left( \Psi, M_2 ( \Psi^2 ) \right) - \dfrac{1}{4} \, \omega_S \left( \Psi, M_3 ( \Psi^3 ) \right) - ... .
\end{equation}
To compute the effective action we fix Siegel gauge for convenience 
\be
b_0\Psi=0,
\ee
and we split the string field into its massless $L_0=0$ component $\Psi_A$ and the remaining massive degrees of freedom $R$,
\be
\Psi=P_0\Psi+\bar P_0\Psi=\Psi_A+R,
\ee
where $P_0$ is the projector on the kernel of $L_0$ and $\bar P_0\equiv 1-P_0$.
 The equation of motion for the massive fields is the projection of the full equation of motion \eqref{eqa} outside the kernel of $L_0$:
\begin{equation}
\bar{P}_0 \left[ Q_B \Psi + M_2( \Psi^2) + M_3( \Psi^3) + ... \right] = 0.
\end{equation}
To solve this equation we expand perturbatively $\Psi$ by means of a small coupling constant $g$
\begin{equation}
\Psi = g \Psi_A + \sum_{n=2}^{+ \infty} g^n R_n.
\end{equation}
We are interested in computing the effective potential of a zero momentum massless field of the form
\be
\Psi_A=c\VV_{\frac12}e^{-\phi},
\ee
which is on-shell and in the small Hilbert space,
\begin{equation}
Q_B \Psi_A = 0 \qquad , \qquad \eta_0 \Psi_A = 0.
\end{equation}
The first non-trivial equation of motion involves $R_2$ and the solution is easily  found
\begin{equation}
\bar{P}_0 \left[ Q_B R_2 + M_2(\Psi_A^2) \right] = 0 \qquad \rightarrow \qquad R_2 = -\dfrac{b_0}{L_0} \, \bar{P}_0 \, M_2 (\Psi_A^2). \label{soleks}
\end{equation}
These are enough to treat the effective action at the fourth order. We recognize that the solution found for the first massive contribution reminds the solution found for the Berkovits theory in \cite{loc1}. Besides the fact that \eqref{soleks} lives in the small Hilbert space, the important difference is the different algebraic structure involved:
\begin{equation}
\left[ \eta_0 \Phi_A, \, Q_B \Phi_A \right] \qquad  \longrightarrow \qquad  M_2(\Psi_A^2),
\end{equation}
where $\Phi_A=\xi_0\Psi_A$.
Since the basic string field is degree even, $R_2$ is degree even too and $Q_B R_2 $ is degree odd. This is consistent with the fact that $M_2(\Psi_A^2)$ is degree odd. The expanded action at the fourth order is then given by
\be
S^{(4)}_{{\rm eff}, A_\infty}(\Psi_A) &=& - \dfrac{1}{2}\, \omega_S \left( \Psi_A, Q_B R_3 \right) - \dfrac{1}{2}\, \omega_S \left( R_3, Q_B \Psi_A \right) - \dfrac{1}{2}\, \omega_S \left( R_2, Q_B R_2 \right) \nonumber \\
&& - \dfrac{1}{3}\, \omega_S \left( R_2, M_2(\Psi_A^2) \right) - \dfrac{1}{3}\, \omega_S \left( \Psi_A, M_2( R_2 , \Psi_A) \right) - \dfrac{1}{3}\, \omega_S \left( \Psi_A, M_2(  \Psi_A , R_2) \right) \nonumber \\
&&  - \dfrac{1}{4} \omega_S \left( \Psi_A, M_3(  \Psi_A^3) \right) ,\0
\ee
which  is easily simplified to
\be
S^{(4)}_{{\rm eff}, A_\infty}(\Psi_A) = + \dfrac{1}{2}\, \omega_S \left( R_2, Q_B R_2 \right) - \dfrac{1}{4} \omega_S \left( \Psi_A, M_3(  \Psi_A^3) \right).
\ee
By substituting $R_2$ this  is a sum of a propagator term and a contact term\footnote{This effective action is proportional to the four-point amplitude derived in  \cite{EKS}, with 4 identical entries.}
\begin{equation}\label{seffeks}
S^{(4)}_{{\rm eff}, A_\infty}(\Psi_A) = - \dfrac{1}{2} \, \omega_S \left(  M_2(\Psi_A^2), \dfrac{b_0}{L_0} \bar{P} M_2(\Psi_A^2) \right) - \dfrac{1}{4} \, \omega_S \left( \Psi_A, M_3 ( \Psi_A^3 ) \right).
\end{equation}
\section{Mapping the $2-$string product}\label{sec:3}

In this section we prepare the ground for proving the equivalence of the effective actions derived from Berkovits and EKS theory. From the classical gauge fixing in the large Hilbert space $\Phi =  \xi_0  \Psi$ we can relate the EKS string field $\Psi$ to the large Hilbert space string field $\Phi$
\begin{equation}
\Psi =  \eta_0 \Phi \qquad \Longleftrightarrow \qquad \Phi =  \xi_0 \Psi. \label{gaugefix}
\end{equation}
Explicitly we consider
\be
\Psi&=&c\VV_{\frac12} e^{-\phi}=c\VV_{\frac12}\delta(\gamma),\label{VV}\\
\Phi&=&c\xi e^{-\phi}\VV_{\frac12}=c\gamma^{-1}\VV_{\frac12}.
\ee
Moreover we can write the BRST variation of the WZW string field as the picture changed $A_\infty$ field
\begin{equation}
Q_B \Phi =  X_0 \Psi.
\end{equation}

With the gauge fixing  \eqref{gaugefix} in mind we can show that the $M_2$ product is related to the commutator  $\left[ \eta_0 \Phi, \, Q_B \Phi \right]$. In terms of the EKS string field, this commutator is written as
\begin{equation}
\left[ \eta_0 \Phi, \, Q_B \Phi \right]=\left[ \Psi, \, X_0 \Psi \right]. \label{berkocom}
\end{equation}
Starting from the definition, it is possible to write the $M_2$ string product in terms of the Berkovits commutator \eqref{berkocom} plus a new composite string field
\begin{eqnarray}
M_2( \Psi, \Psi) &=& \dfrac{1}{3} \left[ X_0 m_2 (\Psi, \Psi) + m_2(X_0 \Psi, \Psi)+ m_2(\Psi, X_0 \Psi)    \right] \nonumber \\
&=& \dfrac{1}{3} \left[ X_0 \Psi^2 + (X_0 \Psi) \Psi + \Psi( X_0 \Psi)    \right] \nonumber \\
&=& \dfrac{1}{3} \left[ X_0 \Psi^2 + \dfrac{3(3-1)}{6}(X_0 \Psi) \Psi + \dfrac{3(3-1)}{6} \Psi( X_0 \Psi)    \right] \nonumber \\
&=& \dfrac{1}{2} \left( (X_0 \Psi) \Psi + \Psi( X_0 \Psi) \right) + \dfrac{1}{3}X_0 \Psi^2  - \dfrac{1}{6} \left( (X_0 \Psi) \Psi + \Psi( X_0 \Psi) \right) \nonumber \\
&=& \dfrac{1}{2} \left[ \Psi, X_0 \Psi \right] + \dfrac{1}{6} Q_B \left( 2 \xi_0 \Psi^2 - \left[\xi_0 \Psi, \Psi \right] \right). 
\end{eqnarray}
We have isolated a Berkovits-like commutator which is in the small Hilbert space, $Q_B$-closed on shell and $Q_B$-exact in the large Hilbert space. The second term is manifestly $Q_B$-exact, and also in the small Hilbert space
\begin{equation}
\eta_0 \, \left( 2 \xi_0 \Psi^2 - \left[\xi_0 \Psi, \Psi \right] \right) =  \left( 2 \Psi^2 - \left[ \Psi, \Psi \right] \right) =  \left( 2  \Psi^2 - 2 \Psi^2  \right) = 0.
\end{equation}
This small Hilbert space string field is the sum of two large Hilbert space string fields. Although these two string fields are different due to the position of $\xi_0$ , they concide at level zero: 
\begin{equation}
P_0 \, \left( 2 \xi_0 \Psi^2 - \left[\xi_0 \Psi, \Psi \right] \right) = 0,
\end{equation}
as it is easy to check. It is also not difficult to check by direct OPE \cite{loc1} that
$$P_0 \left[ \Psi, X_0 \Psi \right] =P_0\left[\eta_0\Phi,Q_B\Phi\right]= 0,$$ 
in total  this means that the following projector condition is satisfied\footnote{This condition has also been discussed recently in \cite{ivo} as the first integrability condition for solutions representing marginal deformations.}
\be
P_0M_2(\Psi,\Psi)=0,
\ee
whenever $\Psi=c\VV_{\frac12} e^{-\phi}$.\\
Defining the Berkovits commutator 
\be
\frac12\left[\Psi, X_0 \Psi \right]\equiv B_2(\Psi, \Psi)
\ee
and
\be
S_2 (\Psi, \Psi)\equiv \dfrac{1}{6}  \left( 2 \xi_0 \Psi^2 - \left[\xi_0 \Psi, \Psi \right] \right),
\ee
 we therefore have
\begin{equation}
M_2 (\Psi, \Psi) = B_2 (\Psi, \Psi) + Q_B S_2 (\Psi, \Psi).
\end{equation}
Working on-shell, from  $B_2$  we can extract a BRST charge
\begin{equation}
B_2 (\Psi, \Psi) = \dfrac{1}{2} \left[ X_0 \Psi,  \Psi \right] = \dfrac{1}{2} Q_B \left[\xi_0 \Psi, \Psi \right]  \equiv Q_B \bar{B}_2 (\Psi, \Psi),
\end{equation}
where in line with the notation of \cite{EKS} we have defined a dressed Berkovits product in the large Hilbert space
\begin{equation}
\bar{B}_2 (\Psi, \Psi) = \dfrac{1}{2} \left[\xi_0 \Psi, \Psi \right].
\end{equation}
So it is natural to define the $A_{\infty}$ dressed product\footnote{The same procedure can be applied to construct another dual string product such that we can extract a $\eta_0$ from $M_2$. However we do not need such products to derive our results.} in terms of $\bar{B}_2, S_2$
\begin{equation}
\bar{M}_2 (\Psi, \Psi) = \bar{B}_2 (\Psi, \Psi) + S_2 ( \Psi, \Psi)
\end{equation}
as a sum of a large Hilbert space commutator and a small Hilbert space contribution lying outside the $L_0 = 0$ subspace. We note also that
\begin{equation}
S_2 (\Psi, \Psi) = \dfrac{1}{6} \, \left( 2\, \xi_0 \, \Psi^2 - 2 \, \bar{B}_2(\Psi, \Psi) \right).
\end{equation}
 
It is easy to see that the above defined products obey the following properties
\begin{equation}
\left[ \eta_0 , M_2 \right] = 0 \qquad , \qquad \left[ \eta_0 , \bar{M}_2 \right] = m_2 
\end{equation}
\begin{equation}
\left[ Q_B , M_2 \right] = 0 \qquad , \qquad \left[ Q_B , \bar{M}_2 \right] = M_2
\end{equation}
\begin{equation}
\left[ Q_B , B_2 \right] = 0 \qquad , \qquad    \left[ Q_B , \bar{B}_2 \right] = B_2 
\end{equation}
\begin{equation}
\left[ \eta_0 , S_2 \right] = 0 \qquad , \qquad \left[ \eta_0 , B_2 \right] = 0
\end{equation}
\begin{equation}
P_0 M_2 = 0 \qquad , \qquad P_0 B_2 = 0 \qquad , \qquad P_0 S_2 = 0.
\end{equation}
The last three relations are true when the products act on a pair of zero momentum on-shell string fields $\Psi$  in the small Hilbert space. We will use these relations extensively in the next sections.

\section{Equality of the effective actions }\label{sec:4}

In this section we prove the equality of the effective actions derived from Berkovits theory and EKS theory, respectively given by\footnote{Our conventions for the SFT trace are $\Tr_S[AB]=\Tr_L[\xi_0AB]=(-1)^{|A|+1}\omega_S(A,B)=(-1)^{Deg(A)}\omega_S(A,B)$ where $|A|$ is the grassmanality}
\be
S^{(4)}_{{\rm eff}, WZW}(\Phi_A)= \dfrac{1}{8} \,\Tr_L\left[[\eta_0 \Phi_A, Q_B\Phi_A]\, \xi_0 \dfrac{b_0}{L_0}\bar{P}_0 \, [\eta_0 \Phi_A,Q_B\Phi_A]\right] -\dfrac{1}{24} \, \Tr_L[[\eta_0 \Phi_A, \Phi_A] \,[\Phi_A, Q_B \Phi_A]], \0 \\
\ee 
\begin{equation}
S^{(4)}_{{\rm eff}, A_\infty}(\Psi_A)= - \dfrac{1}{2} \, \omega_S \left(  M_2(\Psi_A^2), \dfrac{b_0}{L_0} \bar{P}_0 M_2(\Psi_A^2) \right) - \dfrac{1}{4} \, \omega_S \left( \Psi_A, M_3 ( \Psi_A^3 ) \right).
\end{equation}
Since the $A_\infty$ theory is manifestly in the small Hilbert space, we will start by rewriting the WZW theory in the small Hilbert space as well, so that they will  be easier to compare.

\subsection{Berkovits effective action in the small Hilbert space} 

The first step is to write the Berkovits effective action in the small Hilbert space. This is easy for the propagator term which is in the large Hilbert space due to the explicit (and only) $\xi_0$ insertion. Therefore we can simply factorize its one-point function and write everything in the small Hilbert space
\begin{equation}
\Tr_L \left[[\eta_0 \Phi_A,Q_B\Phi_A]\, \xi_0 \dfrac{b_0}{L_0}\bar{P}_0 \, [\eta_0 \Phi_A,Q_B\Phi_A]\right] \equiv \Tr_S \left[[\eta_0 \Phi_A,Q_B\Phi_A]\, \dfrac{b_0}{L_0}\bar{P}_0 \, [\eta_0 \Phi_A,Q_B\Phi_A]\right].
\end{equation}
The same procedure is not that straightforward  for the contact term, which is an honest correlator in the large Hilbert space. Consider the quartic contact vertex in Berkovits theory
\begin{equation}
S_{WZW}^{\rm con}(\Phi_A) = -\dfrac{1}{24}\mathrm{Tr}_L\,[[\eta_0 \Phi_A, \Phi_A] \,[\Phi_A, Q_B \Phi_A]].
\end{equation}
It can be rewritten with the usual gauge fixing $\Phi = \xi_0 \Psi$ as
\begin{equation}
S_{WZW}^{\rm con}(\xi_0 \Psi_A) = \dfrac{1}{24}\mathrm{Tr}_L\,[[\xi_0 \Psi_A, \Psi_A] \,[\xi_0 \Psi_A, X_0\Psi_A]].
\end{equation}
Now we can add a contribution which is identically vanishing
\begin{equation}
\dfrac{1}{24}\mathrm{Tr}_L\,\left[[\xi_0 \Psi_A,  \xi_0 \Psi_A] \,[ \Psi_A, X_0\Psi_A]\right] = 0  \qquad \text{because} \qquad [\xi_0 \Psi_A,  \xi_0 \Psi_A]  = 0.
\end{equation}
Then we have
\begin{eqnarray}
S_{WZW}^{\rm con}(\xi_0 \Psi_A) &=& \dfrac{1}{24}\, \mathrm{Tr}_L\,\left[[\xi_0 \Psi_A, \Psi_A] \,[\xi_0 \Psi_A, X_0\Psi_A] + [\xi_0 \Psi_A,  \xi_0 \Psi_A] \,[ \Psi_A, X_0\Psi_A]\right] \0 \\
&=& \dfrac{1}{24} \, \mathrm{Tr}_L\,\left[\xi_0 \Psi_A   \left( [ \Psi_A, [\xi_0 \Psi_A, X_0\Psi_A] ]+  [ \xi_0 \Psi_A,[ \Psi_A, X_0\Psi_A]]\right) \right]  .\label{sbercon}
\end{eqnarray}
We have collected in round brackets the open string field
\begin{equation}
\Omega = [ \Psi_A, [\xi_0 \Psi_A, X_0\Psi_A] ]+  [ \xi_0 \Psi_A,[ \Psi_A, X_0\Psi_A]],
\end{equation}
which is in the small Hilbert space and can be therefore written as $\eta_0$-exact
\begin{equation}
\Omega = \eta_0 [\xi_0 \Psi_A, [\xi_0 \Psi_A, X_0\Psi_A] ]= \eta_0 \left(ad_{\xi_0 \Psi}^{\,2} X_0 \Psi_A \right).
\end{equation}
 Notice that this is true also off-shell. In terms of this string field we can factorize the one-point function of $\xi_0$ and put the correlator in the small Hilbert space
\begin{equation}
S_{WZW}^{\rm con}(\xi_0 \Psi_A)=S_{WZW}^{{\rm con},\,s}(\Psi_A) = \dfrac{1}{24} \, \mathrm{Tr}_L\,\left[ (\xi_0 \Psi_A) \, \Omega \right]  =  \dfrac{1}{24} \, \mathrm{Tr}_S\,\left[ \Psi_A \, \Omega \right] .
\end{equation}
The Berkovits contact term in the small Hilbert space therefore reads
\begin{equation}
S_{WZW}^{{\rm con},\,s}(\Psi_A) = \dfrac{1}{24} \, \mathrm{Tr}_S\,\left[ \Psi_A   \left( [ \Psi_A, [\xi_0 \Psi_A, X_0\Psi_A] ]+  [ \xi_0 \Psi_A,[ \Psi_A, X_0\Psi_A]]\right) \right]  .
\end{equation}
Taking advantage of the Berkovits products introduced in the previous section, and using a Jacobi identity we can also write the contact term as
\begin{equation}
S_{WZW}^{{\rm con},\,s}(\Psi_A) = \dfrac{1}{12} \, \mathrm{Tr}_S\,\left[ \Psi_A   \left( 2 \, \left[ \xi_0 \Psi_A, B_2(\Psi_A, \Psi_A) \right]  -  \left[ X_0 \Psi_A, \bar{B}_2 (\Psi_A, \Psi_A) \right]\right) \right]  \label{contactsmall}.
\end{equation}
To summarize, the expression of the effective action at quartic order of the WZW theory can be written in the small Hilbert space as
\be
S^{(4)}_{{\rm eff}, WZW}(\xi_0\Psi_A)&=&S^{(4),\,s}_{{\rm eff}, WZW}(\Psi_A)= \frac12\Tr_S \left[B_2(\Psi_A,\Psi_A)\, \dfrac{b_0}{L_0}\bar{P}_0 \,B_2(\Psi_A,\Psi_A)\right]\\
&&+\dfrac{1}{12} \, \mathrm{Tr}_S\,\left[ \Psi_A   \left( 2 \, \left[ \xi_0 \Psi_A, B_2(\Psi_A, \Psi_A) \right]  -  \left[ X_0 \Psi_A, \bar{B}_2 (\Psi_A, \Psi_A) \right]\right) \right].\0
\ee
\subsubsection{WZW action in the small Hilbert space}
As a side remark it should be noticed that once the partial gauge fixing $\Phi=\xi_0\Psi$ is imposed, the whole microscopic WZW action can be written in the small Hilbert space.
To see this consider the  action with the usual gauge fixing
\begin{equation}
S_{WZW}[\xi_0 \Psi] = -\int_{0}^{1}dt\, \text{Tr}_L \left[\Psi \, A_Q (t)\bigl|_{\Phi = \xi_0 \Psi} \right]= -\int_{0}^{1}dt\, \text{Tr}_L \left[\Psi \, A_Q (t)\bigl|_{\Phi = \xi_0 \Psi}\right],
\end{equation}
where we can choose for definiteness
\be
A_Q(t)=e^{-t\Phi}Q(e^{t\Phi}).
\ee
We can always introduce $1 = \left[ \eta_0, \xi_0 \right]$ and find:
\be
S_{WZW}[\xi_0 \Psi] &=& -\int_{0}^{1}dt\, \text{Tr}_L \left[ \left[ \eta_0, \xi_0 \right] \Psi \, A_Q (t)\bigl|_{\Phi = \xi_0 \Psi}\right]\0 \\
&=& -\int_{0}^{1}dt\, \text{Tr}_L \left[  \eta_0 \xi_0 \Psi \, A_Q (t)\bigl|_{\Phi = \xi_0 \Psi}\right]
+ \int_{0}^{1}dt\, \text{Tr}_L \left[   \xi_0 \Psi \, \eta_0 \left( A_Q (t)\bigl|_{\Phi = \xi_0 \Psi} \right)\right].\0 \\
&=& \int_{0}^{1}dt\, \text{Tr}_L \left[   \xi_0 \Psi \, \eta_0 \left( A_Q (t)\bigl|_{\Phi = \xi_0 \Psi} \right)\right],
\ee
where the first term has vanished because $\eta_0$ is a derivation which kills the trace in the large Hilbert space.
Now $\eta_0 (...)$ is clearly in the small Hilbert space. So the remaining trace can be readily rewritten in the small Hilbert space by just dropping  $\xi_0$. Thus in this form the Berkovits action is manifestly in the small Hilbert space
\be
S_{WZW}[\xi_0 \Psi] = \int_{0}^{1}dt\, \text{Tr}_S \left[ \Psi \, \eta_0 \left( A_Q (t)\bigl|_{\Phi = \xi_0 \Psi} \right)\right].\0 \\
\ee

\subsection{Relation between the propagator terms of the $A_\infty$ and WZW effective action}

Now we use the results of the previous section to work on the propagator term in the small Hilbert space.  Hereafter, $\Psi$ denotes the massless component open string field of the effective action. From the previous section, the $M_2$ string product is a sum of the Berkovits fundamental commutator $B_2$ and a term in the small Hilbert space:
\begin{equation}
M_2 (\Psi, \Psi) = B_2 (\Psi, \Psi) + Q_B S_2 (\Psi, \Psi).
\end{equation}
The propagator term 
\begin{equation}
S_{A_\infty}^{\rm prop} (\Psi)= -  \dfrac{1}{2} \, \omega_S \left(  M_2(\Psi^2), \dfrac{b_0}{L_0} \bar{P}_0 M_2(\Psi^2) \right)
\end{equation}
can therefore be written as follows
\begin{eqnarray}
S_{A_\infty}^{\rm prop} (\Psi)&=& - \dfrac{1}{2} \, \omega_S \left(  B_2 (\Psi, \Psi) + Q_B S_2 (\Psi, \Psi), \dfrac{b_0}{L_0} \bar{P}_0 \left( B_2 (\Psi, \Psi) + Q_B S_2 (\Psi, \Psi) \right) \right) \0 \\
&=& + \dfrac{1}{2} \, \Tr_S \left[ \left( B_2 (\Psi, \Psi) + Q_B S_2 (\Psi, \Psi)\right) \dfrac{b_0}{L_0} \bar{P}_0 \left( B_2 (\Psi, \Psi) + Q_B S_2 (\Psi, \Psi) \right) \right] \0 \\
&=& + \dfrac{1}{2} \, \Tr_S \left[ B_2(\Psi, \Psi) \dfrac{b_0}{L_0} \bar{P}_0 \, B_2 (\Psi, \Psi) \right] + \Tr_S \left[ B_2(\Psi, \Psi) \dfrac{b_0}{L_0} \bar{P}_0 \, Q_B S_2 (\Psi, \Psi) \right]\0 \\
& & + \dfrac{1}{2}\, \Tr_S \left[ Q_B S_2(\Psi, \Psi) \dfrac{b_0}{L_0} \bar{P}_0 \, Q_B S_2 (\Psi, \Psi) \right]\0 \\
&=& S_{WZW}^{{\rm prop},\, s}  ( \Psi) + \Upsilon (\Psi).
\end{eqnarray}
In the last line we have isolated the Berkovits propagator term in the small Hilbert space
\begin{equation}
S_{WZW}^{{\rm prop},\, s}  ( \Psi) = \dfrac{1}{2} \, \Tr_S \left[ B_2(\Psi, \Psi) \dfrac{b_0}{L_0} \bar{P}_0 \, B_2 (\Psi, \Psi) \right] = \dfrac{1}{8} \, \Tr_S \left[\left[X_0 \Psi, \Psi \right]\, \dfrac{b_0}{L_0}\bar{P}_0 \, \left[X_0 \Psi, \Psi \right] \right] .
\end{equation}
The extra term  $ \Upsilon (\Psi) $ appearing above is given by 
\begin{eqnarray}
\Upsilon (\Psi) &=&  \Tr_S \left[ \left( B_2 (\Psi, \Psi) + \dfrac{1}{2} Q_B S_2 (\Psi, \Psi) \right) \dfrac{b_0}{L_0} \bar{P}_0 \,  Q_B S_2 (\Psi, \Psi) \right]
\end{eqnarray}
and, despite the appearance, this is not really a propagator term. Recalling that $S_2$ is in the small Hilbert space, we can move the BRST charge to act on its left remaining in the small Hilbert space:
\begin{eqnarray}
\Upsilon (\Psi) &=&  \Tr_S \left[  \left( B_2 (\Psi, \Psi) + \dfrac{1}{2} Q_B S_2 (\Psi, \Psi) \right)\left( \bar{P}_0 - Q_B \dfrac{b_0}{L_0} \bar{P}_0 \right) S_2 (\Psi, \Psi) \right]. \0\\
\end{eqnarray}
Here by using the following identities  which have been demonstrated in the previous section
\begin{equation}
P_0 \, B_2(\Psi, \Psi) = 0 \qquad , \qquad P_0 \, S_2 (\Psi, \Psi) = 0 \qquad , \qquad \left[ Q_B , B_2 (\Psi, \Psi) \right] = 0,
\end{equation}
we can write $\Upsilon (\Psi)$ as a pure contact term without projector components
\begin{eqnarray}
\Upsilon (\Psi) = \Tr_S \left[  \left( B_2 (\Psi, \Psi) + \dfrac{1}{2} Q_B S_2 (\Psi, \Psi) \right)\, S_2 (\Psi, \Psi) \right].
\end{eqnarray}
The trace can be splitted since the string fields $B_2, S_2$ are indipendently in the small Hilbert space. Substituting the explicit expressions for $B_2$, we remain with
\begin{equation}\label{upsilon}
\Upsilon (\Psi) = \dfrac{5}{12}\, \Tr_S \left[ \Psi \,  \left[X_0 \Psi, \, S_2(\Psi, \Psi)\right]\right] + \dfrac{1}{12}\, \Tr_S \left[ \Psi \, \left[\Psi, \, X_0 S_2 (\Psi, \Psi) \right] \right].
\end{equation}

\subsection{Contact terms}

To relate the WZW and $A_\infty$ contact terms we have to work with the $M_3$ string product which for a degree even on-shell string field $\Psi$ is given by 
\begin{eqnarray}
M_3(\Psi, \Psi, \Psi) &=& \dfrac{1}{2} \, M_2(\Psi, \bar{M}_2 (\Psi,\Psi)) + \dfrac{1}{2} \, M_2( \bar{M}_2 (\Psi,\Psi),\Psi) - \dfrac{1}{2} \,\bar{M}_2 (M_2(\Psi,\Psi),\Psi) \nonumber \\
&&  - \dfrac{1}{2} \, \bar{M}_2 (\Psi, M_2(\Psi,\Psi)) +  \dfrac{1}{2} \, Q_B \bar{M}_3\left( \Psi,\, \Psi, \, \Psi \right) .
\end{eqnarray}
We postpone to appendix \ref{app:A} the detailed computation  and here we just quote the result
\begin{equation}
S_{A_\infty}^{\rm con}(\Psi) + \Upsilon (\Psi) = S_{WZW}^{{\rm con},\,s}(\Psi),\label{cont-equa}
\end{equation}
with $\Upsilon$ given in \eqref{upsilon}.
This, together with what we have proven in the previous subsection
\begin{equation}
S_{A_\infty}^{\rm prop}(\Psi) - \Upsilon (\Psi) = S_{WZW}^{{\rm prop},\, s}  ( \Psi) 
\end{equation}
 completes the proof of the equality of the two effective actions. 

\section{Localization in the small Hilbert space}\label{sec:5}

In the previous section we have demonstrated the equivalence of the effective actions derived from WZW and $A_\infty$. Moreover we have shown that the WZW effective action can be written in the small Hilbert space
\begin{equation}
S^{(4)}_{{\rm eff}, A_\infty}(\Psi) = S^{(4)}_{{\rm eff}, WZW}(\xi_0\Psi)=S^{(4),\,s}_{{\rm eff}, WZW}(\Psi) :=S^{(4)}_{\rm eff}(\Psi),
\end{equation}
in the following explicit way
\begin{eqnarray}
S^{(4)}_{\rm eff}(\Psi)&=&  + \dfrac{1}{2} \, \Tr_S \left[ B_2(\Psi, \Psi) \dfrac{b_0}{L_0} \bar{P}_0 \, B_2 (\Psi, \Psi) \right]\0 \\
& &  + \dfrac{1}{12} \, \mathrm{Tr}_S\,\left[ \Psi   \left( 2 \, \left[ \xi_0 \Psi, B_2(\Psi, \Psi) \right]  -  \left[ X_0 \Psi, \bar{B}_2 (\Psi, \Psi) \right]\right) \right].\label{seffreal}
\end{eqnarray}
The non trivial part in the proof of this equality has been to show that no anomalies from the projector on the kernel of $L_0$ arise in this correspondence. This has been ensured by the projector conditions $P_0M_2(\Psi,\Psi)=0$ in the $A_\infty$ theory and the corresponding $P_0[\eta_0\Phi,Q_B\Phi]=0$ in the WZW theory.

In this section we prove (independently of \cite{loc1}) that these equal expressions are fully captured by singular contributions at the boundary of moduli space, which deals with the kernel of $L_0$. 
In \cite{loc1} we have obtained this result working in the large Hilbert space by moving $\eta_0$ and $Q_B$ in the WZW amplitude and thus effectively changing the picture assignements of the entries. Such moves are available in the large Hilbert space  \cite{FMS} but not in the small Hilbert space. To show the localization in the small Hilbert space we have therefore to proceed differently. To this end we  follow Sen's strategy \cite{Sen-restoration}: given a 4-point amplitude, we subtract a companion contribution with a ``wrong'' assignment of the picture.  This extra amplitude is carefully chosen to be zero by some charge/ghost number conservation and therefore it does not change the original result. Since the two amplitudes differ only by an assignement of the picture, their difference is BRST exact in the small Hilbert space. Then we can move the generated BRST charge. When $Q_B$ passes through the propagator it creates a standard contact term in the middle of moduli space (which cancels with the elementary contact term that is already present in the effective action) plus a $P_0$ projector contribution which corresponds to a degenerate four point function at the boundary of moduli space, which ends up giving the whole contribution to the amplitude.
 As in \cite{Sen-restoration} and in \cite{loc1} the needed charge is provided by the $R$-symmetry charge of the $\N=2$ description of the open superstring background we are considering.  \\
\subsection{Conserved charge}

 The needed charge, which we will call $J$, is the  $U(1)$ $R$-symmetry of an
$\mathcal{N}=2$  matter SCFT. The $\N=2$ supersymmetry is global rather than local (as the original $\N$=1) and it is deeply related to the existence of space-time fermions and space-time supersymmetry, see for example \cite{Banks:1988yz,susy2}. 
What typically happens is that the original $\N=1$ supercurrent is given by the sum of the two $\N=2$ supercurrents
\be
T_F=T_F^{(+)}+T_F^{(-)},\label{TFpm}
\ee
which, together with $T(z)$, generate an $\N=2$ superconformal algebra
\be
T(z)\ T(w)&=&\frac{c/2}{(z-w)^4}+\frac{2T(w)}{(z-w)^2}+\frac{\del T(w)}{z-w}+...\\
T(z)\ T_F^{(\pm)}(w)&=&\frac32\frac{ T_F^{\pm}(w)}{(z-w)^2}+\frac{\del T_F^{(\pm)}(w)}{z-w}+...\\
T_F^{(+)}(z)\ T_F^{(-)}(w)&=&\frac{2c/3}{(z-w)^3}+\frac{J(w)}{(z-w)^2}+\frac1{z-w}(2T(w)+\del J(w))+...\\
T(z)\ J(w)&=&\frac{J(w)}{(z-w)^2}+\frac{\del J(w)}{z-w}+...\\
J(z)\ T_F^{(\pm)}(w)&=&\pm\frac{ T_F^{(\pm)}(w)}{z-w}+...\\
J(z)\ J(w)&=&\frac{c/3}{(z-w)^2}+...
\ee
In superstring theory the  matter CFT has $c=15$. However sometimes it is useful to concentrate on subsectors, for example  flat 4-dimensional Minkowski space $(c=6)$ or six dimensional torii or Calabi-Yau compactifications $(c=9)$. All of these very common superstring  backgrounds have a global $\mathcal{N}=2$ superconformal symmetry on their worldsheet.
The case which will be of interest for us is when the $\mathcal{N}=1$ superconformal primary $\VV_{\frac12}$ splits into the sum of two  $\mathcal{N}=2$ superconformal primaries 
\be
\VV_{\frac12}=\VV_{\frac12}^{(+)}+\VV_{\frac12}^{(-)},
\ee
obeying the OPE's
\be
T_F^{(\pm)}(z)\VV^{(\mp)}_{\frac12}(w)&=&\frac1{z-w}\VV^{(\mp)}_{1}(w)+...\label{V1pm}\label{descpm}\\
T_F^{(\mp)}(z)\VV^{(\mp)}_{\frac12}(w)&=&\textrm{regular}.
\ee
The $R$-current $J(z)$  defines a conserved charge
\be
J_0=\oint\frac{dz}{2\pi i}\,J(z),
\ee
and the short superconformal primaries $\VV_{\frac12}^{(\pm)}$ are $J_0$-eigenstates 
\be
J_0 \VV_{\frac12}^{(\pm)}=\pm  \VV_{\frac12}^{(\pm)}.
\ee
From (\ref{TFpm},\ref{V1pm}) we see that the super-descendent matter field $\VV_1$ also decomposes as
\be
\VV_1=\VV_{1}^{(+)}+\VV_{1}^{(-)}.
\ee
However, despite the notation, the super-descendents $\VV_1^{\pm}$ are not charged under $J_0$, because the net $J$-charge in \eqref{descpm} is zero.
In the matter SCFT only correlators with total vanishing $J$-charge are non-zero and this will give a useful selection rule.

From now on we assume that the physical string field $\Psi$ of the $A_\infty$ theory  can be decomposed 
in charged eigenstates of the zero mode  $J_0$ as follows
\begin{equation}
\Psi = \Psi^+ + \Psi^-,\label{split}
\end{equation}
with
\be
J_0\Psi^\pm(z)=\oint_z\frac{dw}{2\pi i} J(w)\Psi^\pm(z)=\pm\Psi^\pm(z).
\ee
Moreover, having in mind that the $\N=2$ susy descendents of $\VV_{\frac12}$ are uncharged, we find that the picture-changed string fields are composed of a charged and uncharged component with different ghost structure
\be
X_0\Psi^{\pm}=X_0 (c\VV_{\frac12}^{(\pm)}\delta(\gamma))=c\VV_{1}^{(\pm)}-\gamma \VV_{\frac12}^{(\pm)}.
\ee
Therefore changing the picture can interfere with $J$ and ghost number conservation and make some amplitude  vanish.
\subsection{Localization of the propagator term} Consider the propagator term of the effective action \eqref{seffreal}. After splitting the physical string field as \eqref{split}, it will decompose in two non-vanishing contributions. These terms are given by
\begin{eqnarray}
S^{\rm prop}(\Psi)&=S_{\pm \pm}^{\rm prop}( \Psi)+S_{\pm \mp}^{\rm prop}( \Psi),
\end{eqnarray}
where
\be
S_{\pm \pm}^{\rm prop}( \Psi) &=& + \dfrac{1}{2} \, \Tr_S \left[ B_2(\Psi^+, \Psi^+) \dfrac{b_0}{L_0} \bar{P}_0 \, B_2 (\Psi^-, \Psi^-) \right]\0 \\
& & + \dfrac{1}{2} \, \Tr_S \left[ B_2(\Psi^-, \Psi^-) \dfrac{b_0}{L_0} \bar{P}_0 \, B_2 (\Psi^+, \Psi^+) \right]\0 \\
&=& + \dfrac{1}{8} \, \text{Tr}_S \left[ \left[ \Psi^+, X_0 \Psi^+ \right] \dfrac{b_0}{L_0} \bar{P}_0  \left[ \Psi^-, X_0 \Psi^- \right] \right]\0\\
& & + \dfrac{1}{8} \, \text{Tr}_S \left[ \left[ \Psi^-, X_0 \Psi^- \right] \dfrac{b_0}{L_0} \bar{P}_0  \left[ \Psi^+, X_0 \Psi^+ \right] \right].
\end{eqnarray}
and 
\be
S_{\pm \mp}^{\rm prop}( \Psi) &=& + \dfrac{1}{2} \, \Tr_S \left[ B_2(\Psi^+, \Psi^-) \dfrac{b_0}{L_0} \bar{P}_0 \, B_2 (\Psi^+, \Psi^-) \right]\0\\
&=& + \dfrac{1}{8} \, \text{Tr}_S \left[ \left[ \Psi^+, X_0 \Psi^- \right]\dfrac{b_0}{L_0} \bar{P}_0  \left[ \Psi^-, X_0 \Psi^+ \right] \right] + (+ \leftrightarrow -). \0\\
\ee
It is clear that  terms obtained by the exchange $(+\leftrightarrow-)$ are equal. However we do not sum them because we want to carry out the calculations in the most symmetric way under the $(+ \leftrightarrow -)$ exchange. 
\subsubsection{$\pm\pm \mp\mp$ Propagator Term}
We start with
\begin{eqnarray}
S_{\pm \pm}^{\rm prop}( \Psi) = + \dfrac{1}{8} \, \text{Tr}_S \left[ \left[ \Psi^+, X_0 \Psi^+ \right] \dfrac{b_0}{L_0} \bar{P}_0  \left[ \Psi^-, X_0 \Psi^- \right] \right] + (+ \leftrightarrow -)
\end{eqnarray}
and we consider the following amplitudes with a different assignement of the picture, which are zero for ghost number and charge conservation
\begin{equation}
\mathcal{A}_1 = \dfrac{1}{8} \Tr_S \left[ \left[ \Psi^+, \Psi^+ \right] \dfrac{b_0}{L_0} \bar{P}_0   \left[ X_0 \Psi^-, X_0 \Psi^- \right] \right] = 0,
\end{equation}
\begin{equation}
\mathcal{A}_2 = \dfrac{1}{8} \Tr_S \left[ \left[ \Psi^-,  \Psi^- \right] \dfrac{b_0}{L_0} \bar{P}_0 \left[ X_0 \Psi^+, X_0 \Psi^+ \right] \right] =\mathcal{A}_1^{(+\leftrightarrow -)}=0 .
\end{equation}
We subtract these vanishing contributions to the original amplitude and get
\begin{eqnarray}\label{pip}
S_{\pm \pm}^{\rm prop} (\Psi) &=& + \dfrac{1}{8} \, \text{Tr}_S \left[  \Psi^+ \left( \left[ X_0 \Psi^+, \dfrac{b_0}{L_0} \bar{P}_0  \left[ \Psi^-, X_0 \Psi^- \right] \right] - \left[ \Psi^+, \dfrac{b_0}{L_0} \bar{P}_0  \left[ X_0 \Psi^-, X_0 \Psi^- \right] \right]\right) \right] \nonumber \\
&& + (+ \leftrightarrow -).
\end{eqnarray}
Now we consider the action of $Q_B$ on the following two string fields in the small Hilbert space, differing only in the assignment of the $R-$charge:
\begin{equation}
\hat{\Psi}_1 = \left[ \xi_0 \Psi^+ , \dfrac{b_0}{L_0} \bar{P}_0  \left[ \Psi^-, X_0 \Psi^- \right]\right]  - \left[ \Psi^+ , \dfrac{b_0}{L_0} \bar{P}_0  \left[ \xi_0 \Psi^-, X_0 \Psi^- \right]\right] \qquad , \qquad \eta_0 \hat{\Psi}_1  = 0, \0
\end{equation}
\begin{equation}
\hat{\Psi}_2 = \left[ \xi_0 \Psi^- , \dfrac{b_0}{L_0} \bar{P}_0  \left[ \Psi^+, X_0 \Psi^+ \right]\right]  - \left[ \Psi^- , \dfrac{b_0}{L_0} \bar{P}_0  \left[ \xi_0 \Psi^+, X_0 \Psi^+ \right]\right] \qquad , \qquad \eta_0 \hat{\Psi}_2  = 0. \0
\end{equation}
Their BRST variation is given by
\begin{eqnarray}
Q_B \hat{\Psi}_1 &=& \left[ X_0 \Psi^+ , \dfrac{b_0}{L_0} \bar{P}_0  \left[ \Psi^-, X_0 \Psi^- \right]\right] - \left[ \Psi^+ , \dfrac{b_0}{L_0} \bar{P}_0 \left[ X_0 \Psi^-, X_0 \Psi^- \right]\right]\0 \\
&& + \left[ \xi_0 \Psi^+ , \bar{P}_0  \left[ \Psi^-, X_0 \Psi^- \right]\right]  + \left[ \Psi^+ , \bar{P}_0  \left[ \xi_0 \Psi^-, X_0 \Psi^- \right]\right],
\end{eqnarray}
\begin{eqnarray}
Q_B \hat{\Psi}_2 &=& \left[ X_0 \Psi^- , \dfrac{b_0}{L_0} \bar{P}_0  \left[ \Psi^+, X_0 \Psi^+ \right]\right] - \left[ \Psi^- , \dfrac{b_0}{L_0} \bar{P}_0  \left[ X_0 \Psi^+, X_0 \Psi^+ \right]\right]\0 \\
&& + \left[ \xi_0 \Psi^- , \bar{P}_0  \left[ \Psi^+, X_0 \Psi^+ \right]\right]  + \left[ \Psi^- , \bar{P}_0  \left[ \xi_0 \Psi^+, X_0 \Psi^+ \right]\right].
\end{eqnarray}
Then the string fields in the small Hilbert space appearing in \eqref{pip}
\begin{eqnarray}
 \left[ X_0 \Psi^+ , \dfrac{b_0}{L_0} \bar{P}_0  \left[ \Psi^-, X_0 \Psi^- \right]\right] - \left[ \Psi^+ , \dfrac{b_0}{L_0} \bar{P}_0  \left[ X_0 \Psi^-, X_0 \Psi^- \right]\right]  \qquad , \qquad  (+ \leftrightarrow -) \0\\
\end{eqnarray}
can be substituted respectively by
\begin{eqnarray}
+ Q_B \hat{\Psi}_1 - \left[ \xi_0 \Psi^+ , \bar{P}_0  \left[ \Psi^-, X_0 \Psi^- \right]\right] - \left[ \Psi^+ , \bar{P}_0  \left[ \xi_0 \Psi^-, X_0 \Psi^- \right]\right], 
\end{eqnarray}
\begin{eqnarray}
 + Q_B \hat{\Psi}_2 - \left[ \xi_0 \Psi^- , \bar{P}_0  \left[ \Psi^+, X_0 \Psi^+ \right]\right] - \left[ \Psi^- , \bar{P}_0  \left[ \xi_0 \Psi^+, X_0 \Psi^+ \right]\right].
\end{eqnarray}
We can deform the contour integral of the BRST charge in the amplitude remaining in the small Hilbert space. The BRST exact state decouples, and we obtain
\begin{eqnarray}
S_{\pm \pm}^{\rm prop}(\Psi) = - \dfrac{1}{8} \text{Tr}_S \left[  \Psi^+ \left[ \left[ \xi_0 \Psi^+ , \bar{P}_0  \left[ \Psi^-, X_0 \Psi^- \right]\right]  + \left[ \Psi^+ ,  \bar{P}_0  \left[ \xi_0 \Psi^-, X_0 \Psi^- \right]\right] \right] \right] + (+ \leftrightarrow -), \0\\ \label{proploc1}
\end{eqnarray}
which is a sum of contact terms and  localized terms. The localized terms  in \eqref{proploc1} are given by
\begin{eqnarray}
S_{\pm \pm}^{P_0} (\Psi) = - \dfrac{1}{8} \text{Tr}_S \left[  \Psi^+ \left[ \left[ \xi_0 \Psi^+ , P_0  \left[ \Psi^-, X_0 \Psi^- \right]\right]  + \left[ \Psi^+ ,  P_0  \left[ \xi_0 \Psi^-, X_0 \Psi^- \right]\right] \right] \right] + (+ \leftrightarrow - ). \0 \\
\end{eqnarray}
These two contributions, just for convenience, can be now computed as a correlator in the large Hilbert space inserting a $\xi_0$ in the amplitude. But in fact we may recognize that the quantity $P_0  \left[ \Psi^-, X_0 \Psi^- \right]=P_0[\eta_0\Phi^{-}, Q_B \Phi^{-}]=0$ is vanishing due to the projector condition in the large Hilbert space. Moreover the quantity $P_0  \left[ \xi_0\Psi^-, X_0 \Psi^- \right]$ is actually in the small Hilbert space thanks to the projection at level zero \cite{loc1} 
\be
P_0  \left[ \xi_0\Psi^-, X_0 \Psi^- \right]=P_0  \left[\Phi^-, Q_B \Phi^- \right]=-2c \HH_1^{(-)}.
\ee
Here the weight 1 primaries $\HH_1^{(\pm)}$ are the charged ``auxialiry fields'' which are obtained by leading order OPE 
\be
\HH_1^{(\pm)}(x)&=&\lim_{\eps\to0}\VV_{\frac12}^{(\pm)}(x+\eps)\VV_{\frac12}^{(\pm)}(x-\eps),
\ee
and have $J$-charge $\pm2$ 
\be
J_0\HH_1^{(\pm)}=\pm2\,\HH_1^{(\pm)}.
\ee
Then we can write
\be
S_{\pm \pm}^{P_0} (\Psi)=- \dfrac{1}{8} \text{Tr}_S {\Big[ }[ \Psi^+ , \Psi^+ ]  P_0  \left[ \xi_0 \Psi^-, X_0 \Psi^- \right] {\Big] } + (+ \leftrightarrow - ).
\ee
A quick comparison shows that this is exactly the same we have found in \cite{loc1} and it can be universally written as a matter two-point function
\be
S_{\pm \pm}^{P_0} (\Psi)=\tr\left[\bra{\HH_{1}^{(+)}}\,\HH_1^{(-)}\rangle\right].
\ee
The above localized contribution is accompanied by an extra contact term in the middle of moduli space which is explicitly given by
\begin{eqnarray}
S_{\pm \pm}^{\rm prop,\, con} (\Psi) =  - \dfrac{1}{8} \text{Tr}_S \left[  \Psi^+ \left[ \left[ \xi_0 \Psi^+ ,  \left[ \Psi^-, X_0 \Psi^- \right]\right]  + \left[ \Psi^+ , \left[ \xi_0 \Psi^-, X_0 \Psi^- \right]\right] \right] \right] + (+ \leftrightarrow -),\0\\ \label{extra1}
\end{eqnarray}
and which will have to cancel against the corresponding contact terms given by the elementary vertices in the effective action.

\subsubsection{$\pm\mp \pm\mp$ Propagator Term}

An analogous computation can be carried out for the second part of the propagator term 
\begin{eqnarray}
S_{\pm \mp}^{\rm prop}( \Psi) &=& + \dfrac{1}{2} \, \Tr_S \left[ B_2(\Psi^+, \Psi^-) \dfrac{b_0}{L_0} \bar{P}_0 \, B_2 (\Psi^+, \Psi^-) \right]+ (+ \leftrightarrow -)\0\\
&=& + \dfrac{1}{8} \, \text{Tr}_S \left[ \left[ \Psi^+, X_0 \Psi^- \right] \dfrac{b_0}{L_0} \bar{P}_0  \left[ \Psi^-, X_0 \Psi^+ \right] \right] + (+ \leftrightarrow -).
\end{eqnarray}
We consider two propagator terms with the wrong assignment of the picture for both the entries, which are zero for ghost and charge conservation
\begin{equation}
\mathcal{A}_3 = \dfrac{1}{8} \, \text{Tr}_S \left[  \left[ \Psi^+, X_0 \Psi^- \right] \dfrac{b_0}{L_0} \bar{P}_0  \left[X_0 \Psi^-, \Psi^+ \right] \right] =0, 
\end{equation}
\begin{equation}
\mathcal{A}_4 =  \dfrac{1}{8} \, \text{Tr}_S \left[ \left[ \Psi^-, X_0 \Psi^+ \right]  \dfrac{b_0}{L_0} \bar{P}_0 \left[ X_0 \Psi^+,  \Psi^- \right]   \right] =0,
\end{equation}
and obtain
\begin{eqnarray}
S_{\pm \mp}^{\rm prop}( \Psi) &=& + \dfrac{1}{8} \, \text{Tr}_S \left[ \Psi^+ \left[  X_0 \Psi^- , \dfrac{b_0}{L_0} \bar{P}_0  \left[ \left[ \Psi^-, X_0 \Psi^+ \right]  - \left[X_0 \Psi^-, \Psi^+ \right] \right] \right]\right] + (+ \leftrightarrow -). \nonumber \\
\end{eqnarray}
We consider the following string field in the small Hilbert space
\begin{equation}
\Psi^0 = - \left[ \left[ \Psi^-, \xi_0 \Psi^+ \right] + \left[\xi_0 \Psi^-,  \Psi^+ \right] \right] \qquad , \qquad 
\eta_0 \Psi^0 = 0,
\end{equation}
which satisfies the following relation
\begin{equation}
Q_B \Psi^0 =  \left[ \Psi^-, X_0 \Psi^+ \right] - \left[X_0 \Psi^-, \Psi^+ \right].
\end{equation}
Then we can extract a BRST charge on the right remaining in the small Hilbert space, and deforming the contour we cancel the propagator. This leaves us  with a projector term at the boundary of moduli space  and a standard contact term. The total neutral  projector term
\begin{eqnarray}
S_{\pm \mp}^{P_0} (\Psi)& =& + \dfrac{1}{8} \, \text{Tr}_S \Bigg{[}\Big{(}\left[\Psi^+, X_0 \Psi^- \right] - \left[ \Psi^-, X_0 \Psi^+ \right]\Big{)} \, P_0 \, \Big{(}\left[ \Psi^-, \xi_0 \Psi^+ \right]  + \left[\xi_0 \Psi^-, \Psi^+ \right] \Big{)} \Bigg{]}\0\\
\end{eqnarray}
can be computed as a correlator in the large Hilbert space inserting a $\xi_0$ in the amplitude. This exactly reproduce the analogous contribution we derived in \cite{loc1} by working entirely in the large Hilbert space. Following \cite{loc1} this can be written as a matter 2-point function
\be
S_{\pm \mp}^{P_0} (\xi_0 \Psi)=\frac14\tr\left[ \bra{\HH_{0}}\,\HH_0\rangle \right],
\ee
where we have defined the neutral auxiliary field $\HH_0$
\be
\HH_0(x)&=&\lim_{\eps\to0}\, (2\eps)\, \left(\VV_{\frac12}^{(+)}(x+\eps)\VV_{\frac12}^{(-)}(x-\eps)-\VV_{\frac12}^{(-)}(x+\eps)\VV_{\frac12}^{(+)}(x-\eps)\right).
\ee
Notice that, differently from the charged pair $\HH^{(\pm)}_1$, this auxiliary field is proportional to the identity vertex operator.
 The above localized contribution comes together with a remaining contact term that will have to cancel against the corresponding elementary vertex in the effective action
 \begin{eqnarray}
S_{\pm \mp}^{\rm prop,\, con} (\Psi) = - \dfrac{1}{8} \, \text{Tr}_S  \left[ \Psi^+ \left[ \left[ X_0 \Psi^- ,  \left[ \Psi^-, \xi_0 \Psi^+ \right]\right] + \left[ X_0 \Psi^- , \left[\xi_0 \Psi^-, \Psi^+ \right] \right] \right] \right] + (+ \leftrightarrow -). \0 \\ \label{extra2}
\end{eqnarray}
Then the extra contact term coming from the propagator term is given by summing \eqref{extra1} and \eqref{extra2}
\begin{equation}
S^{\rm prop,\, con} (\Psi) = S_{\pm \pm}^{\rm prop,\, con} (\Psi) + S_{\pm \mp}^{\rm prop,\, con} (\Psi) .
\end{equation}
\subsection{Elementary vertices} 
When the splitting of the string field is done, the contact terms in the small Hilbert space from the elementary vertices in the effective action \eqref{seffreal} decompose as a sum of six terms
\begin{eqnarray}
S^{\rm con} (\Psi) &=& + \dfrac{1}{24} \, \Tr_S \left[\Psi^+ \left[ \left[ \Psi^+, \left[\xi_0 \Psi^-, X_0 \Psi^- \right] \right] + \left[ \xi_0 \Psi^+, \left[\Psi^-, X_0 \Psi^- \right] \right] \right] \right] \0 \\
& & + \dfrac{1}{24} \, \Tr_S \left[\Psi^+ \left[ \left[ \Psi^-, \left[\xi_0 \Psi^+, X_0 \Psi^- \right] \right] + \left[ \xi_0 \Psi^-, \left[\Psi^+, X_0 \Psi^- \right] \right] \right] \right] \0 \\
& & + \dfrac{1}{24} \, \Tr_S \left[\Psi^+ \left[ \left[ \Psi^-, \left[\xi_0 \Psi^-, X_0 \Psi^+ \right] \right] + \left[ \xi_0 \Psi^-, \left[\Psi^-, X_0 \Psi^+\right] \right] \right] \right] \0 \\
& & +\, (+ \leftrightarrow -).
\end{eqnarray}
We use Jacobi identities
\begin{equation}
 \left[ \Psi^-, \left[\xi_0 \Psi^+, X_0 \Psi^- \right] \right] = \left[ \xi_0 \Psi^+, \left[ X_0 \Psi^-, \Psi^- \right] \right] + \left[ X_0 \Psi^-, \left[\Psi^-, \xi_0 \Psi^+ \right] \right], \label{Jac1}
\end{equation}
\begin{equation}
 \left[ \xi_0 \Psi^-, \left[ \Psi^+, X_0 \Psi^- \right] \right] = \left[ \Psi^+, \left[ \xi_0 \Psi^-, X_0 \Psi^- \right] \right] + \left[ X_0 \Psi^-, \left[ \xi_0 \Psi^-, \Psi^+ \right] \right], \label{Jac2}
\end{equation}
to write 
\begin{eqnarray}
S^{\rm con} (\Psi) &=& + \dfrac{1}{24} \, \Tr_S \left[ \Psi^+ \left[ 2 \left[\Psi^+, \left[\xi_0 \Psi^-, X_0 \Psi^- \right] \right] + 2 \left[\xi_0 \Psi^+, \left[ \Psi^-, X_0 \Psi^- \right] \right] \right. \right. \nonumber\\
& & \left. \left. \quad\quad\quad\quad\quad\quad+  \left[ X_0 \Psi^-, \left[ \Psi^-,\xi_0 \Psi^+ \right] \right] + \left[ X_0 \Psi^-, \left[ \xi_0 \Psi^-, \Psi^+ \right] \right] \right. \right. \nonumber \\
& & \left. \left.  \quad\quad\quad\quad\quad\quad+ \left[\Psi^-, \left[\xi_0 \Psi^-, X_0 \Psi^+ \right] \right] + \left[\xi_0 \Psi^-, \left[ \Psi^-, X_0 \Psi^+ \right] \right] \right] \right]\0\\ 
& &+ \, (+ \leftrightarrow -). \label{contactloc}
\end{eqnarray}
\subsection{Cancelation of contact terms}
The first two lines of the contact term \eqref{contactloc} are already in the required form to simplify the extra contact terms \eqref{extra1} and \eqref{extra2} coming from the propagator term. We focus our attention to the last line. To manipulate these terms we need to keep all the $(+ \leftrightarrow -)$ terms together:
\begin{eqnarray}
S^{{\rm con},\,3} (\Psi) &=& + \dfrac{1}{24} \, \Tr_S \left[ \Psi^+ \left[ \left[\Psi^-, \left[\xi_0 \Psi^-, X_0 \Psi^+ \right] \right] + \left[\xi_0 \Psi^-, \left[ \Psi^-, X_0 \Psi^+ \right] \right] \right] \right]\0\\
& & + \dfrac{1}{24} \, \Tr_S \left[ \Psi^- \left[ \left[\Psi^+, \left[\xi_0 \Psi^+, X_0 \Psi^- \right] \right] + \left[\xi_0 \Psi^+, \left[ \Psi^+, X_0 \Psi^- \right] \right] \right] \right].
\end{eqnarray}
It is convenient to put these terms in the large Hilbert space and rearrange the trace. The trace $\Tr_L$ in the large Hilbert space is defined simply as
\begin{equation}
\Tr_L \left[ \xi_0 (...) \right] = \Tr_S \left[ (...) \right].
\end{equation}
Once that we are in the large Hilbert space we can split the traces and use the cyclicity of the commutators to change the string field in the first position. In particular we move in the first position the $\Psi^{-}$ string field in the first line and the $\Psi^{+}$ string field in the second line. We obtain
\be
S^{{\rm con},\,3} (\Psi) &=& - \dfrac{1}{24} \, \Tr_L \left[ \Psi^- \left[ \left[  X_0 \Psi^+, \left[\xi_0 \Psi^-, \xi_0 \Psi^+ \right] \right] +  \left[\xi_0 \Psi^+, \left[ \xi_0 \Psi^-, X_0 \Psi^+ \right] \right] \right] \right]\0 \\
& & - \dfrac{1}{24} \, \Tr_L \left[ \Psi^+ \left[ \left[  X_0 \Psi^-, \left[\xi_0 \Psi^+, \xi_0 \Psi^- \right] \right] +  \left[\xi_0 \Psi^-, \left[ \xi_0 \Psi^+, X_0 \Psi^- \right] \right] \right] \right]\0 \\
&=& - \dfrac{1}{24} \, \Tr_L \left[ \Psi^+ \left[ \left[  X_0 \Psi^-, \left[\xi_0 \Psi^+, \xi_0 \Psi^- \right] \right] +  \left[\xi_0 \Psi^-, \left[ \xi_0 \Psi^+, X_0 \Psi^- \right] \right] \right] \right] \0 \\
& & + (+ \leftrightarrow -).
\ee
The results of this manipulation can be put back in the small Hilbert space reabsorbing $\xi_0$ as follows. We insert $1 = \left[\eta_0 , \xi_0 \right]$ first,
\be
S^{{\rm con},\,3} (\Psi) &=& - \dfrac{1}{24} \, \Tr_L \left[ \left( \left[\eta_0, \xi_0 \right]  \Psi^+ \right) \left[ \left[  X_0 \Psi^-, \left[\xi_0 \Psi^+, \xi_0 \Psi^- \right] \right] +  \left[\xi_0 \Psi^-, \left[ \xi_0 \Psi^+, X_0 \Psi^- \right] \right] \right] \right] \0 \\
& & +  (+ \leftrightarrow -)\0\\
&=& - \dfrac{1}{24} \, \Tr_L \left[ \eta_0 \xi_0   \Psi^+  \left[ \left[  X_0 \Psi^-, \left[\xi_0 \Psi^+, \xi_0 \Psi^- \right] \right] +  \left[\xi_0 \Psi^-, \left[ \xi_0 \Psi^+, X_0 \Psi^- \right] \right] \right] \right] \0 \\
& & +  (+ \leftrightarrow -)\0\\
&=& + \dfrac{1}{24} \, \Tr_L \left[ \left( \xi_0 \Psi^+ \right)  \eta_0 \left[ \left[  X_0 \Psi^-, \left[\xi_0 \Psi^+, \xi_0 \Psi^- \right] \right] +  \left[\xi_0 \Psi^-, \left[ \xi_0 \Psi^+, X_0 \Psi^- \right] \right] \right] \right] \0 \\
& & +  (+ \leftrightarrow -).
\ee
Now we have a first entry in the large Hilbert while the rest is manifestly in the small Hilbert space. Thus we remain with
\begin{eqnarray}
S^{{\rm con},\,3} (\Psi) &=& + \dfrac{1}{24} \, \Tr_S \left[ \Psi^+  \left[  \left[X_0 \Psi^-, \left[ \xi_0 \Psi^-,  \Psi^+ \right] \right] + \left[X_0 \Psi^-, \left[ \Psi^-, \xi_0 \Psi^+ \right] \right] \right]  \right] \nonumber\\
& & + \dfrac{1}{24} \, \Tr_S \left[ \Psi^+  \left[   \left[ \Psi^-, \left[\xi_0 \Psi^+, X_0 \Psi^- \right] \right] + \left[\xi_0 \Psi^-, \left[ \Psi^+, X_0 \Psi^- \right] \right] \right] \right]  \0\\
& &  + (+ \leftrightarrow -).
\end{eqnarray}
A further application of Jacobi identities \eqref{Jac1}, \eqref{Jac2} to the second line of the last equation leads to 
\begin{eqnarray}
S^{{\rm con}, \, 3} (\Psi) &=& + \dfrac{1}{24} \text{Tr}_S \left[  \Psi^+ \left[ \left[ \xi_0 \Psi^+ ,  \left[ \Psi^-, X_0 \Psi^- \right]\right]  + \left[ \Psi^+ , \left[ \xi_0 \Psi^-, X_0 \Psi^- \right]\right] \right] \right]\0 \\
& & + \dfrac{2}{24} \, \text{Tr}_S  \left[ \Psi^+ \left[ \left[ X_0 \Psi^- ,  \left[ \Psi^-, \xi_0 \Psi^+ \right]\right] + \left[ X_0 \Psi^- , \left[\xi_0 \Psi^-, \Psi^+ \right] \right] \right] \right] \0 \\
& & + (+ \leftrightarrow -). 
\end{eqnarray}
Summing the results, we obtain that
\begin{eqnarray}
S^{\rm con} (\Psi) &=& + \dfrac{3}{24} \text{Tr}_S \left[  \Psi^+ \left[ \left[ \xi_0 \Psi^+ ,  \left[ \Psi^-, X_0 \Psi^- \right]\right]  + \left[ \Psi^+ , \left[ \xi_0 \Psi^-, X_0 \Psi^- \right]\right] \right] \right]\0 \\
& & + \dfrac{3}{24} \, \text{Tr}_S  \left[ \Psi^+ \left[ \left[ X_0 \Psi^- ,  \left[ \Psi^-, \xi_0 \Psi^+ \right]\right] + \left[ X_0 \Psi^- , \left[\xi_0 \Psi^-, \Psi^+ \right] \right] \right] \right] \0 \\
& & + (+ \leftrightarrow -), 
\end{eqnarray}
and then
\begin{equation}
S^{\rm con} (\Psi)+ S^{\rm prop,\, con} (\Psi) = 0.
\end{equation}
Therefore all the contact terms at the middle of moduli space cancel and we are left with a fully localized effective action, the same in the $A_\infty$ and WZW theories
\be
S^{(4)}_{\rm eff}(\Psi_A)&=&- \dfrac{1}{8} \text{Tr}_S {\Bigg[ }[ \Psi^+ , \Psi^+ ]  P_0  \left[ \xi_0 \Psi^-, X_0 \Psi^- \right] {\Bigg] }
- \dfrac{1}{8} \text{Tr}_S {\Bigg[ }[ \Psi^- , \Psi^- ]  P_0  \left[ \xi_0 \Psi^+, X_0 \Psi^+ \right] {\Bigg] }\0\\
&&+ \dfrac{1}{8} \, \text{Tr}_S \Bigg{[}\Big{(}\left[\Psi^+, X_0 \Psi^- \right] - \left[ \Psi^-, X_0 \Psi^+ \right]\Big{)} \, P_0 \, \Big{(}\left[ \Psi^-, \xi_0 \Psi^+ \right]  + \left[\xi_0 \Psi^-, \Psi^+ \right] \Big{)} \Bigg{]}\0\\
&=&\tr\left[\bra{\HH_{1}^{(+)}}\,\HH_1^{(-)}\rangle + \frac14\bra{\HH_{0}}\,\HH_0\rangle \right].
\ee
\section{Flat directions and generalized ADHM constraints}\label{sec:6}
The exact universal form of the effective quartic potential of the open superstring we have found 
\be
S_{\rm eff}^{(4)}=\tr\left[\bra{\HH_{1}^{(+)}}\,\HH_1^{(-)}\rangle + \frac14\bra{\HH_{0}}\,\HH_0\rangle \right],
\ee
can be a convenient starting point for a systematic search for flat directions in the full open superstring potential or, in other words, for the search of exact solutions for marginal deformations of non-trivial $D$-brane systems. Although not all of the marginal directions  we might be interested in fall into the $\N=2$ representations we are assuming, many interesting cases are captured by this scheme. 
In full generality we immediately see that a sufficient condition for having a vanishing quartic potential is to set to zero the auxiliary fields
\be
\HH_1^{(\pm)}&=&0,\\
\HH_0&=&0.
\ee
By the very definition of the $\HH$ fields \eqref{H1}, \eqref{H2}, these three equations will in general give quadratic constraints on the zero momentum space-time polarizations of the physical fields $\Psi^{(\pm)}$ and in this regards can be considered as generalized ADHM constraints \cite{ADHM}.
To be concrete on this interpretation, in \cite{loc1} we have analysed the exact quartic potential for the $D3/D(-1)$ system  \cite{Witten:1995im, Douglas:1996uz, torinesi} by using our localization method.  Ignoring transverse degrees of freedom to the $D3$ branes,  the physical zero momentum fields living on the $D3/D(-1)$ system can be assembled into a matrix string field of the form
\begin{equation}
\Psi_A (z)=c  \VV_\frac{1}{2}\, e^{-\phi}(z)= -ce^{-\phi} \left(\begin{matrix}
A & \omega \\
\bar{\omega} & a
\end{matrix}\right) (z),
\end{equation}
where (see \cite{loc1} for details)
\begin{equation}
A(z)=A_{\mu} \psi^{\mu}(z),
\end{equation}
\begin{equation}
\omega(z) = \omega^{N \times k}_{\alpha}  \, \Delta S^{\alpha}(z),
\end{equation}
\begin{equation}
\bar{\omega}(z) = \bar{\omega}^{k \times N}_{\alpha} \, \bar{\Delta} S^{\alpha}(z),
\end{equation}
\begin{equation}
a(z)=a_{\mu}  \psi^{\mu}(z).
\end{equation}
The integers $N$ and $k$ refer respectively to the number of coincident $D3$ branes and the number of $D(-1)$ branes which are initially sitting on the $D3$ worldvolume. 
The ${\cal N}=2$ structure is not manifest in the covariant four-dimensional language and, as common, we have to pass to complex variables on $\mathbb{C}^2$.
We refer to \cite{loc1} for the details of the decomposition and the computation and here we just remind that in this case the auxiliary fields take the form
\be
\HH_{1}^{\left( +\right)} = -\dfrac{i}{4} \, \eta_{-}^{\mu  \nu } \, T_{\mu  \nu} \, \psi_{1}\psi_{ 2}  \vert 0 \rangle, \,
\ee
\be
\HH_{1}^{\left( -\right)} = +\dfrac{i}{4} \, \eta_{+}^{\mu \nu } \, T_{\mu  \nu} \, \psi_{\bar1}\psi_{\bar 2} \vert 0 \rangle, \,
\ee
\be
\HH_{0} = -\dfrac{i}{2} \, \eta_{3}^{\mu  \nu } \, T_{\mu  \nu} \, \vert 0 \rangle, \,
\ee
where we have defined the complex  combinations of the 't Hooft symbols
\be
\eta_{+}^{\mu \nu } \equiv \eta_1^{\mu \nu} +i \eta_2^{\mu \nu} \qquad , \qquad  \eta_{-}^{\mu \nu } \equiv \eta_1^{\mu \nu} -i \eta_2^{\mu \nu}.
\ee
The covariant tensor  $T^{\mu  \nu}$ is given by
\be
T^{\mu  \nu} = \left(\begin{matrix}
 \left[  A^{\mu} ,A^{\nu} \right]  + \dfrac{1}{2} \, \omega_{\alpha} \left( \gamma^{ \mu \nu}\right) ^{\alpha \beta} \bar{\omega}_{\beta} & 0 \\
0 & \left[  a^{\mu} ,a^{\nu} \right]  - \dfrac{1}{2} \, \bar{\omega}_{\alpha} \left( \gamma^{ \mu \nu}\right) ^{\alpha \beta} \omega_{\beta} 
\end{matrix}\right).\label{Tmunu}
\ee
The effective quartic potential thus takes the form
\be
S^{(4)}_{\rm eff}[A,a,w,\bar w] = \tr\left[\bra{\HH_{1}^{(+)}}\,\HH_1^{(-)}\rangle+\frac14\bra{\HH_{0}}\,\HH_0\rangle \right]= -\dfrac{1}{16}\, \tr \left[D_a D_a\right] ,\label{eff-inst}
\ee
where 
\be
D_a=\eta_a^{\mu\nu}\, T_{\mu\nu}.\label{gen-ADHM}
\ee
It is interesting to see whether it is possible to set the $D_a$ to zero and thus to find a flat potential (to order 4) for this system.
Setting $D_a=0$ implies two sets of equations, respectively living on the $D(-1)$ and on the $D3$
\be
\eta^c_{\mu\nu} \left( \left[  a^{\mu} ,a^{\nu} \right]  - \dfrac{1}{2} \, \bar{\omega}_{\alpha} \left( \gamma^{ \mu \nu}\right) ^{\alpha \beta} \omega_{\beta}\right)&=&0\label{ADHM1}\\
\eta^c_{\mu\nu} \left( \left[  A^{\mu} ,A^{\nu} \right]  + \dfrac{1}{2} \, \omega_{\alpha} \left( \gamma^{ \mu \nu}\right) ^{\alpha \beta} \bar{\omega}_{\beta}  \right)&=&0.\label{ADHM2}
\ee
The first set of equations on the $D(-1)$ are the well known three ADHM constraints. Solving them should give a VEV to the fields $a^\mu$ and $w,\bar w$. The second set of three equations is not familiar in the study of gauge theory instantons but it appears quite clear that they will imply a VEV for the zero momentum gauge field on the D-brane, to compensate for the switching on of $w, \bar w$.
Just as the ADHM constraints these set of equations are not easy to solve, but in the case of $k=1$ and (just to stay minimal) $N=2$ we can provide a concrete solution. We start solving  for \eqref{ADHM1} for $k=1$ (where the term $[a^\mu,a^\nu]=0$) at a given size modulus $\rho^2=\bar w^{\alpha} w_{\alpha}$
\be
w_{\alpha}^{\quad i}&=&\frac\rho{\sqrt{ 2}} \left(\begin{matrix}1&0\\0&1\end{matrix}\right)\\
\bar w^{\alpha\,i}&=&\frac\rho{\sqrt{ 2}} \left(\begin{matrix}1&0\\0&1\end{matrix}\right),
\ee
where $\alpha=1,2$ (spanning the row) is the $SU(2)$ spin index and $i=1,2$ (spanning the columns) is the $U(2)$ color index.
Using the explicit form of the 't Hooft symbols and some standard $SO(4)$ spinor/gamma algebra (conveniently summarized in the appendix of \cite{loc1} or \cite{torinesi})
we find an explicit solution of \eqref{ADHM2} in the form
\be
A_\mu=\frac{\rho}{2}\,\sigma_\mu=\frac{\rho}{2}\, (1,-i \vec\tau).
\ee
Therefore we  find that the quartic potential we compute is consistent with the existence of exactly marginal deformations corresponding to blowing up moduli of $D(-1)$'s inside the $D3$ and that, to such order, the possible obstructions are precisely avoided by the three  generalized ADHM constraints which in turn originate from the three different localization channels of the effective action.

\section{Conclusions}\label{concl}
In this paper we have considered the computation of the effective action for massless zero momentum fields to quartic order in the framework of the $A_\infty$ open supestring field theory in the small Hilbert space. In particular we have shown that, up to this order, there is no difference between the effective actions in the WZW or the $A_\infty$ theory. While this may be expected from the fact that the two theories are equivalent upon partial gauge fixing and field redefinition, it is still non trivial, because at zero momentum we generically encounter subtle  contributions at the boundary of moduli space that can invalidate formal manipulations. Nonetheless we find that these potential anomalies are zero thanks to the fundamental projector condition \eqref{proj-cond} and the two effective actions are indeed identical to 4th order. It would be interesting to see if the WZW and $A_\infty$ effective actions will start deviating at higher order due to some higher order non-vanishing contribution at the boundary of moduli space.

If the massless fields we are interested in are charged under an $\N=2$ $R$-symmetry  $J_0$, then the quartic potential fully localizes at the boundary of moduli space and becomes computable in terms of two-point functions of the so-called auxiliary fields. These auxiliary fields are obtained by leading order OPE of the physical fields and they represent the effect of the localization of the four point amplitude to the boundary.
Our localization mechanism, first uncovered in the WZW theory \cite{loc1} and here confirmed in the $A_\infty$ theory, is quite powerful and it would be very interesting to extend it to higher orders in the effective action. Another interesting direction is to study similar localization mechanisms in closed superstring field theories in order to by-pass the explicit construction of four and possibly higher-points couplings, which are far from the boundary of moduli space where we expect the effective action to localize. Given the fact that the $\N=2$ structure is generic in superstring theory we expect that similar localization mechanisms should be available for both Heterotic and type II superstrings. We hope that this research direction will turn useful  for the development of superstring field theory.
  
\section*{Acknowledgments}

We thank Ted Erler and Ashoke Sen for enlighting discussions and suggestions during  various stages of this work. We also thank Marco Bill\'o, Luca Mattiello, Igor Pesando, Rodolfo Russo, Ivo Sachs, Martin Schnabl and in particular Jakub Vosmera for recent discussions.
The authors thank the organizers of the workshop ``Discussion Meeting on String Field Theory'' at HRI, Allahabad, February 2018 where this work initiated. CM thanks  the organizers of the workshop ``New Frontiers in String Theory''  in Kyoto, July 2018 for kind hospitality. 
CM thanks the Galileo Galilei Institute for having hosted the workshop ``String Theory from a worldsheet perspective" spring 2019,  which provided an invaluable 
environment.
This work is partially supported by the MIUR PRIN
Contract 2015MP2CX4 ``Non-perturbative Aspects Of Gauge Theories And Strings''.

\appendix 
\section{Contact terms in the equality of the effective actions}\label{app:A}
In this appendix we show the computations needed to obtain the relation between the elementary quartic vertices of the WZW and the $A_\infty$ theory, shown in the main text \eqref{cont-equa}.
We recall here the definitions of the necessary ingredients to rewrite this product in terms of elementary $B_2, S_2, \bar{B}_2$ string products. The bare and dressed multi-string products for a degree even on-shell string field $\Psi$ are turned in graded Lie algebra commutators as follows
\begin{equation}
M_2 (\Psi, \Psi) = B_2 (\Psi, \Psi) + Q_B S_2 (\Psi, \Psi),
\end{equation} 
\begin{equation}
\bar{M}_2 (\Psi, \Psi) = \bar{B}_2 (\Psi, \Psi) + S_2 (\Psi, \Psi),
\end{equation} 
\begin{equation}
m_3(\Psi, \Psi, \Psi) = \dfrac{2}{3} \, \left[ \Psi, \, \xi_0 \left(\Psi^2 \right) \right],
\end{equation}
\begin{equation}
\bar{M}_3 \left( \Psi, \Psi, \Psi \right) = \dfrac{1}{6} \, \left[ \xi_0 \left[ \Psi, \, \xi_0 \left( \Psi^2 \right) \right]  + \left[\xi_0 \Psi, \, \xi_0 \left( \Psi^2 \right) \right]  + \left[ \Psi, \, \xi_0 \left[ \xi_0  \Psi , \Psi\right]  \right]  \right].
\end{equation}
Moreover we need also
\begin{eqnarray}
M_2(\Psi, \bar{M}_2 (\Psi,\Psi)) + M_2( \bar{M}_2 (\Psi,\Psi),\Psi) &=& \dfrac{1}{3} \left[ X_0 \left[\Psi, \,\bar{B}_2 (\Psi, \Psi) + S_2 (\Psi, \Psi)\right] \right. \nonumber\\
&& + \left. \left[ X_0  \Psi, \, \bar{B}_2 (\Psi, \Psi) + S_2 (\Psi, \Psi)\right] \right. \nonumber \\
&& + \left. \left[ \Psi, \, X_0 \bar{B}_2 (\Psi, \Psi) + X_0 S_2 (\Psi, \Psi)\right]\right], \0 \\
\end{eqnarray}
\begin{eqnarray}
\bar{M}_2 (M_2(\Psi,\Psi),\Psi) +  \bar{M}_2 (\Psi, M_2(\Psi,\Psi)) &=& \dfrac{1}{3} \left[ \xi_0 \left[\Psi, \, B_2 (\Psi, \Psi) + Q_B S_2 (\Psi, \Psi)\right] \right. \nonumber\\
&& + \left. \left[ \xi_0  \Psi, \, B_2 (\Psi, \Psi) + Q_B S_2 (\Psi, \Psi)\right] \right. \nonumber \\
&&  \left. - \left[ \Psi, \, \xi_0 B_2 (\Psi, \Psi) + \xi_0 Q_B S_2 (\Psi, \Psi)\right]\right], \0 \\
\end{eqnarray}
and
\begin{eqnarray}
Q_B \bar{M}_3\left( \Psi, \Psi, \Psi \right) &=& \dfrac{1}{3} \left[ \frac{1}{2} \, X_0 \left[\Psi, \, \bar{B}_2 (\Psi, \Psi) + 3 S_2 (\Psi, \Psi)\right] \right. \0 \\
&& \left. +  \dfrac{1}{2} \, \left[ X_0 \Psi, \, \bar{B}_2 (\Psi, \Psi) + 3 S_2 (\Psi, \Psi)\right] \right. \nonumber\\
&& + \left. \dfrac{1}{2} \, \xi_0 \left[  \Psi, \, B_2 (\Psi, \Psi) + 3 Q_B S_2 (\Psi, \Psi)\right] \right. \nonumber \\
&& + \left. \dfrac{1}{2} \,  \left[ \xi_0 \Psi, \, B_2 (\Psi, \Psi) + 3 Q_B S_2 (\Psi, \Psi)\right] \right. \nonumber \\
&& - \left. \left[ \Psi, \, X_0 \bar{B}_2 (\Psi, \Psi)\right] + \left[ \Psi, \xi_0 B_2 (\Psi, \Psi)\right] \right].
\end{eqnarray}
In terms of these products we can write the full contact term as
\begin{equation}
S_{A_\infty}^{\rm con}(\Psi) =-\frac12\omega_S\left(\Psi,M_3(\Psi^3)\right)= - \dfrac{1}{24} \omega_S \left( \Psi, \chi \right),
\end{equation}
where we have introduced the string field $\chi$ in the small Hilbert space
which is given by
\begin{eqnarray}
\chi &=& X_0 \left[ \Psi, \, \dfrac{3}{2} \bar{B}_2 (\Psi, \Psi) + \dfrac{5}{2} S_2 (\Psi, \Psi)\right] + \left[ X_0 \Psi, \, \dfrac{3}{2} \bar{B}_2 (\Psi, \Psi) + \dfrac{5}{2} S_2 (\Psi, \Psi) \right]  \0 \\
&& + \xi_0 \left[  \Psi, - \dfrac{1}{2} \, B_2 (\Psi, \Psi) + \dfrac{1}{2} Q_B S_2 (\Psi, \Psi)\right] +  \left[ \xi_0 \Psi, - \dfrac{1}{2} \, B_2 (\Psi, \Psi) + \dfrac{1}{2} Q_B S_2 (\Psi, \Psi)\right] \nonumber \\
&& + \left[ \Psi, \, 2\, \xi_0 B_2 (\Psi, \Psi) + X_0 S_2 (\Psi, \Psi) + \xi_0 Q_B S_2(\Psi, \Psi) \right] .
\end{eqnarray}
In order to check that $\eta_0 \chi = 0$, it is necessary to use a Jacobi identity
\begin{equation}
\left[ X_0 \Psi, \left[\Psi, \Psi \right] \right] = - 2 \left[ \Psi, \left[ X_0 \Psi, \Psi \right] \right].
\end{equation}
The total contribution to the contact term which must be compared to the Berkovits contact term \eqref{contactsmall} is then given by
\begin{eqnarray}
S_{A_\infty}^{\rm con}(\Psi) + \Upsilon (\Psi) = - \dfrac{1}{24} \, \omega_S \left( \Psi, \, \chi' \right) = - \dfrac{1}{24} \Tr_S \left[ \Psi \, \chi' \right] , \label{chi'}
\end{eqnarray}
where $\Upsilon(\Psi)$ is defined in \eqref{upsilon} and
\begin{eqnarray}
\chi' &=& X_0 \left[ \Psi, \, \dfrac{3}{2} \bar{B}_2 (\Psi, \Psi) + \dfrac{5}{2} S_2 (\Psi, \Psi)\right] + \left[ X_0 \Psi, \, \dfrac{3}{2} \bar{B}_2 (\Psi, \Psi) - \dfrac{15}{2} S_2 (\Psi, \Psi) \right]  \0 \\
&& + \xi_0 \left[  \Psi, - \dfrac{1}{2} \, B_2 (\Psi, \Psi) + \dfrac{1}{2}\, Q_B S_2 (\Psi, \Psi)\right] +  \left[ \xi_0 \Psi, - \dfrac{1}{2} \, B_2 (\Psi, \Psi) + \dfrac{1}{2} \, Q_B S_2 (\Psi, \Psi)\right] \nonumber \\
&& + \left[ \Psi, \, 2\, \xi_0 B_2 (\Psi, \Psi) - Q_B \xi_0 S_2(\Psi, \Psi) \right] \label{ki} 
\end{eqnarray}
is another string field in the small Hilbert space. Until now we have carried out all the computations in the small Hilbert space. Now for simplicity, we evaluate the correlator \eqref{chi'} in the Large Hilbert space, where we can take advantage of the following identities
\begin{equation}
\Tr_L \left[ (\xi_0 A) \, (\xi_0 B )\right] = 0,
\end{equation}
\begin{equation}
\Tr_L \left[ A  \,(X_0 B) \right] =  \Tr_L  \left[ (X_0 A)\, B \right].
\end{equation}
Evaluating the correlator in the large Hilbert space allow us to get rid of the first term in the second line in  \eqref{ki}. The final result in the Large Hilbert space is given by a sum of five Witten traces. We also substitute the definitions of the $B_2, \bar{B}_2$ and $S_2$ string fields. Then we have:
\begin{eqnarray}
S_{A_\infty}^{\rm con}(\Psi) + \Upsilon (\Psi) &=& + \dfrac{3}{4} \, S_{WZW}^{{\rm con},\, s} (\Psi) - \dfrac{3}{4 \cdot 24} \Tr_L \left[ \left[\xi_0 X_0 \Psi, \Psi \right] \,  \left[\xi_0 \Psi, \Psi \right] \right] \0 \\
& & - \dfrac{5}{6 \cdot 24} \Tr_L \left[ \left[\xi_0 \Psi, X_0 \Psi \right] \,  \left[\xi_0 \Psi, \Psi \right] \right] \0 \\
& & - \dfrac{5}{6 \cdot 24} \Tr_L \left[ \left[\xi_0 \Psi, \Psi \right] \, \xi_0  \left[ \Psi,  X_0 \Psi \right] \right] \0\\
& & + \dfrac{5}{6 \cdot 24} \Tr_L \left[ \left[\xi_0 \Psi, X_0 \Psi \right] \, \xi_0 \eta_0 \left[\xi_0 \Psi, \Psi \right] \right].
\end{eqnarray}
where we have already isolated the Berkovits contact term \eqref{contactsmall}. To simplify the result, we note that:
\begin{itemize}
\item The sum of the last three terms is zero. Indeed
\begin{eqnarray}
\Tr_L \left[ \left[\xi_0 \Psi, X_0 \Psi \right] \, \xi_0 \eta_0 \left[\xi_0 \Psi, \Psi \right] \right] &=& + \Tr_L \left[ \left[\xi_0 \Psi, X_0 \Psi \right] \, \left[\xi_0 \Psi, \Psi \right] \right] \0\\
& & - \Tr_L \left[ \left[ \Psi, X_0 \Psi \right] \, \xi_0  \left[\xi_0 \Psi, \Psi \right] \right],
\end{eqnarray}
so that summing the last three lines we get a vanishing result.
\item We use a Jacobi identity 
\begin{equation}
\left[ \Psi, \left[ \xi_0 \Psi, \Psi \right] \right] = \dfrac{1}{2} \, \left[ \xi_0 \Psi, \left[ \Psi, \Psi \right] \right]
\end{equation}
on the second term. This gives
\begin{eqnarray}
- \dfrac{3}{4 \cdot 24} \Tr_L \left[ \left[\xi_0 X_0 \Psi, \Psi \right] \,  \left[\xi_0 \Psi, \Psi \right] \right] &=& + \dfrac{3}{8 \cdot 24} \Tr_L \left[ \left[X_0 \Psi, \xi_0 \Psi \right] \,  \left[\xi_0 \Psi, \Psi \right] \right]\0\\
& & + \dfrac{3}{8 \cdot 24}\Tr_L \left[ \left[\xi_0 X_0 \Psi, \Psi \right] \,  \left[\xi_0 \Psi, \Psi \right] \right]. \0\\
\end{eqnarray}
Then we find that
\begin{equation}
- \dfrac{3}{4 \cdot 24} \Tr_L \left[ \left[\xi_0 X_0 \Psi, \Psi \right] \,  \left[\xi_0 \Psi, \Psi \right] \right] = + \dfrac{1}{4} \, S_{WZW}^{\rm con} (\xi_0 \Psi).
\end{equation}
\end{itemize}
Therefore we have obtained
\begin{equation}
S_{A_\infty}^{\rm con}(\Psi) + \Upsilon (\Psi) = S_{WZW}^{{\rm con},\, s}(\Psi) .
\end{equation}

%

\section{Comments on  recent approach to instanton marginal deformations}\label{app:B}
In \cite{ivo} it is reported that the equation of motion for the instanton marginal deformation we have  discussed in section \ref{sec:6} is obstructed at order 3. Here, without adding 
 new computations, we would like to make contact with our results.
Writing the solution perturbatively
\be
\Psi=\sum_{n=1}^\infty g^n \Psi_n,\label{sol}
\ee
the equation of motion will take the generic recursive form
\be
Q_B\Psi_n=-EOM_n(\Psi_{n-1},...,\Psi_1),
\ee
where $EOM_n$ is what one gets by varying the interaction terms of the $A_\infty$-action \eqref{Ainf-act} and expanding according to \eqref{sol}.
The first order solution $\Psi_1$ coincides with the physical field $\Psi_A$ \eqref{PsiA}, of which we have computed the effective action.

 To start with, we observe that the effective action at quartic order \eqref{seffeks} can be written as
\be
S_{{\rm eff},\,A_\infty}^{(4)}(\Psi_1)&=& - \dfrac{1}{2} \, \omega_S \left(  M_2(\Psi_1^2), \dfrac{b_0}{L_0} M_2(\Psi_1^2) \right) - \dfrac{1}{4} \, \omega_S \left( \Psi_1, M_3 ( \Psi_1^3 ) \right)\0\\
&=& - \frac14 \, \omega_S \left( \Psi_1, \left[ -M_2\left(\Psi_1, \dfrac{b_0}{L_0}  M_2(\Psi_1^2)  \right)-M_2\left( \dfrac{b_0}{L_0}  M_2(\Psi_1^2) ,\Psi_1 \right)+M_3 ( \Psi_1^3 )\right]\right)\0\\
&=& - \frac14\omega_S\left(\Psi_1,EOM_3[\Psi_1]\right) = - \frac14\omega_S(\Psi_1,P_0 EOM_3[\Psi_1]),\label{seff3}
\ee
where we have used 
that $P_0\Psi_1=\Psi_1$. Here $EOM_3[\Psi_1]$ means that we have solved already $\Psi_2$ in terms of $\Psi_1$  $$Q_B\Psi_2+EOM_2(\Psi_1)=Q_B\Psi_2+M_2(\Psi_1,\Psi_1)=0$$ with
$$\Psi_2=-\frac{b_0}{L_0}M_2(\Psi_1,\Psi_1)$$ and plugged it inside $EOM_3(\Psi_2,\Psi_1)$
\be
EOM_3[\Psi_1]&:=&EOM_3\left(-\frac{b_0}{L_0}M_2(\Psi_1,\Psi_1),\Psi_1\right)\\
&=&-M_2\left(\Psi_1, \dfrac{b_0}{L_0}  M_2(\Psi_1^2)  \right)-M_2\left( \dfrac{b_0}{L_0}  M_2(\Psi_1^2) ,\Psi_1 \right)+M_3 ( \Psi_1^3 ).\0
\ee
If the effective action at quartic order \eqref{seff3} vanishes this means that $EOM_3[\Psi_1]$  does not have components along 
the first order solution $\Psi_1=\Psi_A$.  From our  discussion in section \ref{sec:6} this happens when the generalized ADHM constraints \eqref{ADHM1}, \eqref{ADHM2} are implemented.
As also discussed in \cite{ivo}, the integrability condition for the equation of motion at third order 
\be
Q_B\Psi_3=-EOM_3[\Psi_1]
\ee
is 
\be
P_0\, EOM_3[\Psi_1]=0.
\ee
If this is satisfied then we automatically have that $S_{\rm eff}^{(4)}=0$ because $\Psi_1=P_0\Psi_1$, \eqref{seff3}.
However in \cite{ivo} it is reported that given a generic state in the kernel of $L_0$, which we may call $\chi(=P_0\chi)$, they find $\omega_S\left(\chi,EOM_3[\Psi_1]\right)\neq0$.\\
This is not consistent with our results.
To see this choose $\chi=\Psi_1$. Then $P_0\,  EOM_3[\Psi_1]$ contracted with the first order solution $\Psi_1$ should just give the effective action at quartic order by \eqref{seff3}. It is then easy to see that \cite{ivo} disagrees from us by the absence of 't Hooft symbols as it can be checked by comparing for example eq (7.22) in (ver. 2 of) \cite{ivo} (with $B^{\mu}\to A^{\mu}$ and $v\to w$) with \eqref{eff-inst}. In other words, the effective action \eqref{seff3} computed according to \cite{ivo} would be $$S_{\rm eff}^{(4)}\sim\tr[T_{\mu\nu}T^{\mu\nu}]\,\quad {\rm (incorrect)}$$ with $T_{\mu\nu}$ given in \eqref{Tmunu}, while we find instead $$S_{\rm eff}^{(4)}\sim\tr[(T_{\mu\nu}\eta^{\mu\nu}_c)(\eta^c_{\rho\sigma} T^{\rho\sigma})],$$
consistently with other works on the subject (see $e.g.$ \cite{torinesi, Dorey:2002ik}). 

Without 't Hooft symbols the constraints \eqref{ADHM1} and \eqref{ADHM2} would not be solvable because they would contain too many equations (six instead of three) and the potential could not be made flat.
But in fact there are only three constraints as a natural consequence of the  localization mechanism we have presented, which involves three auxiliary fields and which is in turn based on the underlying $\N=2$ SCFT structure. 
We hope these comments will be useful to settle the issue and to progress in our understanding of $D$-branes moduli in superstring field theory.

\end{document}